# Role of the interplay between spinodal decomposition and crystal growth in the morphological evolution of crystalline bulk heterojunctions

*Olivier J.J. Ronsin [a]\* and Jens Harting [a,b]\**

[a] O. Ronsin, Prof. J. Harting, Helmholtz Institute Erlangen-Nürnberg for Renewable Energy, Forschungszentrum Jülich, Fürther Straße 248, 90429 Nürnberg, Germany
E-mail: o.ronsin@fz-juelich.de, j.harting@fz-juelich.de

[b] Prof. J. Harting, Department of Applied Physics, Eindhoven University of Technology, PO box 513, 5600MB Eindhoven, The Netherlands

**Keywords**: organic solar cells, stability, phase-field, crystallization, liquid-liquid demixing

# Abstract

The stability of organic solar cells is strongly affected by the morphology of the photoactive layers, whose separated crystalline and/or amorphous phases are kinetically quenched far from their thermodynamic equilibrium during the production process. The evolution of these structures during the lifetime of the cell remains poorly understood. In this paper, a phase-field simulation framework is proposed, handling liquid-liquid demixing and polycrystalline growth at the same time in order to investigate the evolution of crystalline immiscible binary systems. We find that initially, the nuclei trigger the spinodal decomposition, while the growing crystals quench the phase coarsening in the amorphous mixture. Conversely, the separated liquid phases guide the crystal growth along the domains of high concentration. It is also demonstrated that with a higher crystallization rate, in the final morphology, single crystals are more structured and form percolating pathways for each material with smaller lateral dimensions.

# Introduction

Organic solar cells (OSC) are a promising technology in the field of photovoltaics. They can be solution-processed onto flexible substrates, opening the way to straightforward, eco-friendly and low-cost manufacturing, as well as novel applications such as flexible or semi-transparent solar modules. Even if their performance stays still far behind the performance of classical silicon solar cells, the best efficiencies of OSC on the laboratory scale has rapidly increased from 10% in 2014 [1] and 13%-14% in 2017 [2,3] to more than 16% in 2019 [4-6]. The efficiency of large area organic solar modules also quickly increases and recently passed 11%. [7] Nevertheless, stability is a second important challenge the organic photovoltaics (OPV) community is facing on the way to commercial products. Whereas silicon solar cells are inherently very stable and have a lifetime of 20 years and more, the performance of OSC tends to drop faster with time. The state of the art is that typical lifetimes of 2-4 years can be currently reached [8] with the best reported extrapolated lifetime being about 10 years. [9] Despite these encouraging results, the stability of OSC still needs to be improved.

The instability of OSC is related to various different extrinsic and intrinsic degradation mechanisms.[8,10,11] Chemical degradation of the electrodes, interfaces or photoactive layers due to UV light or the reaction with water and oxygen are known as extrinsic degradation mechanisms and can be strongly mitigated by using appropriate but expensive encapsulation methods. [12,13] Intrinsic degradation mechanisms are due to the temperature field, as well as visible and IR light absorption inside the solar cell and can lead to the formation of charge blocking layers at the interface or to the chemical evolution of the photoactive layer (PAL) itself. For instance, photo-dimerization of PCBM ([6,6]-phenyl-$C_{61}$-butyric acid methyl ester) has been shown to lead to a strong loss of performance due to the reduction of the electron mobility in polymer-fullerene solar cells.[14]

Morphological degradation of the PAL might also be a significant source of intrinsic instability. In general, photoactive layers of OSC are multi-phase bulk-heterojunction structures made out of at least one donor and one acceptor material. The importance of the bulk-heterojunction (BHJ) morphology for the photovoltaic performance has been identified to be crucial for the OSC performance for a long time. [1,15-24] In these structures, the excitons generated through light absorption are strongly bound and have to be separated into free charge carriers at interfaces between both materials. Since the mean free path of an exciton is typically 10nm, this length scale should also be characteristic for the size of the phases. Moreover, percolated pathways to the electrodes should be available in order to efficiently extract the charge carriers. Finally, relatively pure donor and acceptor crystal phases are desired to ensure high charge carrier mobility. As a consequence, the PAL is typically composed of crystal phases of the donor and acceptor materials and an amorphous mixed phase [25,26] like in the well-known system made of P3HT (poly(3-hexylthiophene)) and PCBM. [27-29]

However, these morphologies are typically partially quenched far from the thermodynamic equilibrium during the deposition process and hence not stable. Therefore, they can evolve during post-processing or operation of the solar cell. Since the sizes and topology of the various phases in the PAL are theoretically expected to influence the optoelectronic properties, this should have an impact on the performance: the stability of the device and stability of the photoactive layer morphology are related. Even if to the best of our knowledge, a direct relationship between structure and optoelectronic performance has not been positively proven, structure-property correlations have been proposed [20,22,30]. P3HT-PCBM films containing micrometer-scaled crystals are known to show very poor efficiencies. More recently, for PCE11 (PffBT4T-2OD) – PCBM systems, demixing of donor and acceptor in the amorphous phase and subsequent aggregation of fullerenes has been proposed as the mechanism for strong burn-in degradation, on the basis of coupled GISAXS and GIWAXS experiments. [31] Different successful strategies have been then proposed to overcome this problem, such as adding a second acceptor, more compatible with the donor in order to stabilize the mixed phase [32-35], the use of non-fullerene acceptors like IDTBR (rhodanine-benzothiadiazole-coupled indacenodithiophene) [36] or to kinetically quench the mixed phase. [37]

These improvements are based on remarkable efforts to unravel the mechanisms of the PAL morphology formation and stability. On the one hand, they base on thermodynamic considerations such as miscibility of solvent, donor and acceptors [32,34,38,39] and the evaluation of phase diagrams [23-24,26,40,41]. On the other hand, the importance of the kinetic evolution towards the thermodynamic equilibrium during the deposition process and ageing, including possible transient or stable liquid-liquid phase separation (LLPS) [42-45], has been acknowledged and recently qualitatively taken into account for stability improvement [37]. Nevertheless, no general coherent physical framework has been proposed to understand the BHJ morphology formation and stability taking into account at the same time thermodynamic aspects such as liquid/amorphous phase stability and crystallinity and the kinetics of the system (kinetics + thermodynamics, crystallinity + miscibility). Therefore, the understanding of stability is still very material system-dependent.

In the last decade, different simulation approaches have been proposed to contribute to the understanding of the BHJ formation and evolution in OPV systems. At the molecular scale, coarse-grained molecular dynamics (CGMD) [46-48], dissipative particle dynamics (DPD) [49-51] or self-consistent field theory (SCFT) [52-53] studies have been performed. However, the length scales and predominately timescales accessible with such techniques prevent them from being used for a kinetic modelling of the whole evolution process, though simulations on about (100nm)$^3$ volumes may be performed.

To tackle this problem and deal with kinetic aspects of structure evolution, the system has therefore to be described at a larger scale within the framework of continuum mechanics. In that context, phase-field simulations can be relevant: the phase-field method is a well-established continuum-mechanics diffuse interface simulation framework for solving the kinetic evolution of thermodynamically complex systems with many phases. [54-57] On the one hand, it has been widely used to investigate the crystallization in single-material systems (metals [58] or polymers [59]) or many-material systems (alloys [60-62], precipitation from a solution [63-64]). Thereby, many crystal systems can be simulated with the multiple-field phase-field (MFPF) [54-55] or with the orientation-field phase field (OFPF) [65-68] approaches. On the other hand, it has been applied to understand the onset and kinetics of LLPS in immiscible binary or ternary fluid mixtures. [56,69-76]. In the field of organic electronics, phase-field modelling has been used to study spinodal decomposition of the mixtures during the drying of the wet film but remains limited to a few papers. Wodo and co-workers dealt extensively with ternary systems including a polymer, a fullerene and an evaporating solvent, adding also specific interactions with the substrate [77,78]. They gained insight into the impact of process parameters on the final amorphous two-phase structure. Michels and co-workers investigated donor/acceptor mixtures with an evaporating solvent [79-83]. With respect to stability, Ray, Alam and co-workers performed simulations of binary amorphous immiscible acceptor-donor systems based on the Cahn-Hilliard equation. They aimed at describing the impact of annealing on phase separation. [84-86]. Finally, both processes of LLPS and crystallization can be coupled in a very natural way within a phase-field framework. However, very few groups have been dealing with both at the same time. Zhou investigated numerically spinodal induced crystallization, showing that the phase separation could promote crystal nucleation and growth, [87] Rathi investigated the competition between crystallization and LLPS in a crystalline-amorphous binary system, [88] and Saylor and Kim simulated the evaporation of an immiscible amorphous-crystalline system in solution with application to polymer-embedded drug-release films. [89-91] In these papers however, only one of the materials is able to crystallize, and no specific interaction between crystals prevents them from merging. Analytical investigations of the interplay between crystallization and LLPS are also limited. For example, Mitra calculated the rate of nucleation for heterogeneous nucleation generated at the interfaces of a coarsening phase-separated liquid mixture. [92] A possible issue with continuum mechanics simulations (for example using the phase-field method) is that the typical grid resolution and interface thicknesses, which depend on the material properties, often turn out to be about 1-3nm and down to 5-15nm, respectively. This is also the case in the present paper. For material systems containing polymers, whereby the Kuhn length of the chain can be of a few nanometers (typically 5nm for P3HT [93]) and the radius of gyration about 10-20nm, the continuum mechanics assumptions might be hardly fulfilled on the scale of the grid resolution. Therefore, the very local structure obtained with phase-field simulations at these length scales for polymer systems should be considered cautiously. Nevertheless, the mesh size and interface thickness obtained in phase-field simulations often rely on rather crude material parameter estimates, whereas they should be carefully measured for each material system. Only under these conditions, the comparison with the size of the molecules and the discussion on the validity of the continuum mechanics assumptions can be conclusive. Additionally, the successful application of phase-field simulations in the past decades, even in the field of polymer science, shows that larger simulated structures (beyond 50nm in scale) are at least qualitatively correct, because the evolution equations on these scales are correctly recovered.

In this paper, we propose a new phase-field model which takes into account the miscibility of the liquid materials as well as their respective crystallization properties. As a consequence, it can handle crystallization of each material and LLPS at the same time. The impingement of single crystals is also included, so that polycrystalline structures can be investigated. Based on simulations of simple binary model systems, we illustrate how it can be used to investigate the stability of OPV photoactive layers, independent of the materials being crystalline or amorphous, miscible or immiscible in the amorphous phase. Through these examples, we outline interaction mechanisms between LLPS and crystallization, and show how the kinetic properties of the system can strongly affect the transient and final morphology of the bulk-heterojunction.

## Model equations

### The free energy functional

A phase-field framework is used to simulate the kinetic evolution of the system towards its thermodynamic equilibrium. The system is composed of $n$ materials, out of which $n_{cryst}$ can have crystal phases. We describe the system morphology with the respective volume fractions $\varphi_i$ of all materials in the system, but also with an order parameter $\Phi_k$ for each crystalline material, whose value varies between 0 in the amorphous phase and 1 in the crystal phase. Additionally, for each crystalline material, orientation parameter fields describe the orientation of each single crystal. For 2D simulations, one single orientation field $\theta_k$ per material is sufficient, whereby two or three angles would be necessary in 3D depending on the crystal symmetry. The 3D case is not discussed in the following, but the generalization is straightforward. Each crystal has its own orientation $[-\pi; \pi]$ and is assumed to remain constant during the simulation, while no orientation is defined in the amorphous phases. The thermodynamic properties of the system are defined with the help of the free energy functional:

$$G_{tot} = \int_V \left(\Delta G_V^{loc} + \Delta G_V^{nonloc}\right) dV, \qquad (1)$$

where $V$ is the total volume. $\Delta G_V^{loc}$ is the local free energy density and $\Delta G_V^{nonloc}$ the non-local contribution due to the field gradients. The local part of the free energy is given by

$$\Delta G_V^{loc}(\{\varphi_i\}, \{\Phi_k\}) = \begin{array}{l} \Delta G_V^{ideal}(\{\varphi_i\}) + \Delta G_V^{inter}(\{\varphi_i\}, \{\Phi_k\}) \\ + \Delta G_V^{SL}(\{\varphi_i\}, \{\Phi_k\}) + \Delta G_V^{num}(\{\varphi_i\}) \end{array}. \qquad (2)$$

The first term on the right-hand side of the equation above represents the free energy density change upon ideal mixing,

$$\Delta G_V^{ideal}(\{\varphi_i\}) = \frac{RT}{v_0} \sum_{i=1}^n \frac{\varphi_i \ln \varphi_i}{N_i}, \qquad (3)$$

with $R$ being the gas constant and $T$ the temperature. Following the Flory-Huggins theory [94], $v_0$ is the molar volume of the lattice site considered to calculate the free energy of mixing. The molar volume of the fluid $i$ is $v_i = N_i v_0$ and $N_i$ its molar size of in terms of lattice units. The interactions between materials are represented by the second term:

$$\Delta G_V^{inter}(\{\varphi_i\}, \{\Phi_k\}) = \frac{RT}{v_0} \begin{pmatrix} \sum_{i=1}^n \sum_{j>i}^n \varphi_i \varphi_j \chi_{ij,ll} \\ + \sum_{k=1}^{n_{cryst}} \sum_{j \neq k}^n \Phi_k^2 \varphi_k \varphi_j \chi_{kj,sl} \\ + \sum_{k=1}^{n_{cryst}} \sum_{j \neq k}^{n_{cryst}} \Phi_j \Phi_k \varphi_k \varphi_j \chi_{kj,ss} \end{pmatrix}, \qquad (4)$$

Matkar and Kyu proposed an extension of the Flory-Huggins theory for binary systems with crystalline materials. [95-96] Equation *(4)* is simply a generalization of their theory for any number of materials. The first term is the classical Flory-Huggins interaction term where $\chi_{ij,ll}$ is the interaction parameter between the amorphous phases of materials $i$ and $j$. Now, the crystalline materials might have a crystal phase which interact with amorphous phases, especially in the diffuse solid-liquid interface. The second term stands for these interactions, with $\chi_{kj,sl}$ representing the interaction between the amorphous phase of material $j$ and the solid phase of material $k$. This term can be understood considering that $\Phi_k$ can be interpreted as the proportion of material $k$ being crystallized, so that $\varphi_k \Phi_k$ is the quantity of solid and $\varphi_j \Phi_k$ the amount of amorphous phase interacting with this solid. [96] The last term stands for the solid-solid interactions, with $\chi_{kj,ss}$ being the interaction parameter between the solid phases of the materials $k$ and $j$. Similar to Matkar and Kyu, we write $\chi_{kj,ss} = c\sqrt{\chi_{kj,sl}}\sqrt{\chi_{jk,sl}}$ with the coefficient $c$ ranging from *-2* for fully compatible crystals to *0* for fully incompatible crystals. Note that we also tested a slightly different form of the interaction terms in Equation *(4)*, with the amorphous- amorphous interactions written as $(1-\Phi_k)(1-\Phi_j)\varphi_k\varphi_j\chi_{kj,ll}$, the solid-liquid interaction terms as $\Phi_k(1-\Phi_j)\varphi_k\varphi_j\chi_{kj,sl}$ and the solid-solid interaction terms as $\Phi_k\Phi_j\varphi_k\varphi_j\chi_{kj,ss}$. Comparison between both forms showed that the interface profiles turn out to be slightly different. However, no major differences were seen in the simulation results and we performed all simulations discussed in this paper using Equation *(4)*.

The third term on the RHS of Equation *(2)* stands for the free energy density of phase change, according to what is commonly used for the simulation of crystallization in metals and alloys [54-55],

$$\Delta G_V^{SL}(\{\varphi_i\},\{\Phi_k\}) = \sum_{k=1}^{n_{cryst}} \rho_k\varphi_k\big(g(\Phi_k)H_k + p(\Phi_k)\Delta G_{V,k}^{cryst}\big). \tag{5}$$

In the equation above, $\rho_k$ is the density of the material *k* and $\Delta G_{V,k}^{cryst} = L_k\left(\dfrac{T}{T_{m,k}}-1\right)$ its free energy density of crystallization, whereby $L_k$ and $T_{m,k}$ are its enthalpy of fusion and melting temperature, respectively. If $\Delta G_{V,k}^{cryst} < 0$, the free energy of the crystal phase is smaller than that of the amorphous phase and the material *k* is prone to crystallize. There is an energy barrier in the solid-liquid phase transition when $\Phi_k$ varies from 0 to 1, provided that $\left|\dfrac{3\Delta G_{V,k}^{cryst}}{H_k}\right| < 1$. The height of the barrier is determined by the parameter $H_k$. $p(\Phi_k)$ and $g(\Phi_k)$ are the interpolation functions:

$$\begin{cases} g(\Phi_k) = \Phi_k^2(\Phi_k-\xi_{0,k})^2 \\ p(\Phi_k) = \Phi_k^2(3\xi_{0,k}-2\Phi_k) \end{cases} \tag{6}$$

Note that other functional forms can be used with no considerable impact on the model behavior. $\xi_{0,k}$ is the value of the order parameter for which the free energy density of phase change is minimized and can be seen as the crystallinity of the material, $\xi_{0,k}=1$ representing a fully crystalline material. As a consequence, semi-crystalline materials can also be considered with such a model.

The fourth term on the RHS of Equation *(2)* is a purely numerical contribution meant to facilitate the convergence properties of the simulation: for common and physically relevant parameter sets (for instance high $N\chi$ values for a highly immiscible amorphous binary polymer system), the expected equilibrium volume fractions in the separated phases are very close to 0 and 1, so that unrealistically small time steps have to be used for the calculation to converge. To overcome this problem, a contribution to the free energy is added if the volume fractions approach 0 and 1. We choose the following form:

$$\Delta G_V^{num}(\{\varphi_i\}) = \begin{cases} k_{num}(\varphi_{num}-\varphi_i)^{n_{num}} & if \quad \varphi_i < \varphi_{num} \\ k_{num}(\varphi_i-(1-\varphi_{num}))^{n_{num}} & if \quad 1-\varphi_{num} < \varphi_i \\ 0 & elsewhere \end{cases} \tag{7}$$

Hence, this numeric correction term has no impact on the properties of the system provided $1-\varphi_{num} < \varphi_i < \varphi_{num}$. The parameter $\varphi_{num}$ has to be kept small in order to minimize the impact on the phase diagram of the model, especially on the volume fractions of the separated phases. For very pure phases not fulfilling this condition, the phase composition in the simulation will deviate from the physical behavior but will remain below $\varphi_{num}$ or above $1 - \varphi_{num}$. $\varphi_{num}$ can be regarded as the numerical precision for the composition of very pure phases. $n_{num}$ and $k_{num}$ are arbitrary numeric parameters fixing the intensity of the numeric correction term.

Finally, the non-local part of the free energy functional describes the contributions of the concentration gradients and the solid-liquid phase change to the surface tension as:

$$\Delta G_V^{nonloc}(\{\nabla\varphi_i\},\{\Phi_k\},\{\theta_k\}) = \begin{aligned}&\sum_{i=1}^{n}\frac{\kappa_i}{2}(\nabla\varphi_i)^2 + \sum_{i=1}^{n_{cryst}}\frac{\varepsilon_k^2}{2}(\nabla\Phi_k)^2\\&+\sum_{i=1}^{n_{cryst}}p(\Phi_k)\left(\varepsilon_{g1,k}|\nabla\theta_k| + \frac{\varepsilon_{g2,k}^2}{2}|\nabla\theta_k|^2\right)\end{aligned}, \qquad (8)$$

where $\kappa_i$ is the surface tension parameter for the concentration gradient of material $i$, $\varepsilon_k$ are the surface tension parameters for the gradient of the order parameter of material $k$, and $\varepsilon_{g1,k}$ and $\varepsilon_{g2,k}$ are the surface tension parameters for the orientation gradients of material $k$. The last term of Equation (8) stands for the orientation mismatch energy between different single crystals of a given material and is responsible for impingement of the crystallites provided the orientation mismatch is sufficiently large. This functional form has been proposed in previous work during the development of the OFPF model, and the reader is referred to the corresponding papers for the description of the properties of the calculated grain boundaries. [65-67] The surface tension between two amorphous phases $i$ and $j$ is proportional to $\sqrt{\kappa_i + \kappa_j}$, whereas the surface tension of a solid-liquid interface also contains a contribution from the phase variation and from the orientation mismatch when two grains impinge. However, the surface tension also depends on other thermodynamic properties such as the molar volumes and interaction parameters and can be computed with standard methods described elsewhere [55,56].

## Kinetic equations

Since they are conserved quantities, the volume fractions obeys the celebrated Cahn-Hilliard equation, initially proposed by Cahn and Hilliard for binary mixtures [97,98] and generalized later for multicomponent mixtures [71-74,80,82,99]:

$$\frac{\partial\varphi_i}{\partial t} = \frac{v_0}{RT}\nabla\left[\sum_{j=1}^{n-1}\Lambda_{ij}\nabla\left(\mu_{V,j}^{gen} - \mu_{V,n}^{gen}\right)\right] \qquad i = 1\ldots n-1 \qquad (9)$$

This is actually a set of coupled continuity equations, where the material fluxes are proportional the driving force for the system evolution, namely the gradient of the exchange chemical potential. In Equation (9), $\mu_{V,j}^{gen}$ is the chemical potential density, defined as the functional derivative of the free energy functional:

$$\mu_{V,j}^{gen} - \mu_{V,n}^{gen} = \left(\frac{\partial \Delta G_V}{\partial \varphi_j}\right) - \left(\frac{\partial \Delta G_V}{\partial \varphi_n}\right) - \left(\nabla\left(\frac{\partial \Delta G_V}{\partial (\nabla \varphi_j)}\right) - \nabla\left(\frac{\partial \Delta G_V}{\partial (\nabla \varphi_n)}\right)\right) \quad (10)$$

The first two terms stands for to the local chemical potential, whereas the two last contributions take into account the potential due to concentration gradients and hence to surface area variations. The symmetric Onsager mobility coefficients $\Lambda_{ij} = \Lambda_{ji}$ have to depend not only on the diffusion coefficients but also on the local mixture composition in order to ensure the incompressibility constraint and the Gibbs-Duhem relationship. Several theories have been proposed to derive correct expressions for the flux, among which the "slow mode theory" [100] and the "fast-mode theory" [101] are the most successful. Their names come from the fact that the mutual diffusion coefficient in a binary system is controlled by the slowest component in the "slow-mode theory", while it is controlled by the fastest component in the "fast-mode theory". The controversy between both theories is not fully resolved yet. However, the fast-mode theory seems to better match experimental data and can also be derived from the general Maxwell-Stefan equations framework [102] in a consistent way. Relying on these arguments, we choose to use the fast-mode theory in this paper, which leads to the following expression of the mobility coefficients:

$$\begin{cases} \Lambda_{ii} = (1-\varphi_i)^2 \omega_i + \varphi_i^2 \sum_{k=1, k\neq i}^{n} \omega_k \\ \Lambda_{ij} = -(1-\varphi_i)\varphi_j \omega_i - (1-\varphi_j)\varphi_i \omega_j + \varphi_i \varphi_j \sum_{k=1, k\neq i\neq j}^{n} \omega_k \end{cases} \quad (11)$$

Here, the coefficients $\omega_i$ are related to the self-diffusion coefficients $D_{s,i}$ of the materials $i$ through:

$$\omega_i = \frac{v_0}{RT} N_i \varphi_i D_{s,i}(\{\varphi_i\}, \{\phi_k\}). \quad (12)$$

The self-diffusion coefficients themselves are also dependent on the mixture composition $\{\varphi_i\}$, but unless specified otherwise, they are kept constant in this paper for simplicity. However, diffusion processes are expected to be dramatically slower in the solid crystal phases. This is accounted for with a hyperbolic tangent based dependence of the diffusion coefficients on the order parameters:

$$D_{s,i}(\{\varphi_i\}, \{\phi_k\}) = D_{s,i}^{liq}\left[1 + \frac{1}{2}\left(tanh(-k_D \phi_D) - tanh(-k_D(\phi_{tot} - \phi_D))\right)\right] \quad (13)$$

Here, $D_{s,i}^{liq}$ is the diffusion coefficient of the material $i$ in the amorphous phases, $\phi_D$ is the value of the order parameter around which the mobility drop is centered, $k_D$ controls the intensity and the steepness of the diffusion coefficient gradient from the amorphous phase to the solid phase, and $\phi_{tot}$ is an estimate of the overall crystallinity at a given position calculated as

$$\phi_{tot} = 1 - \prod_{k=1}^{n_{cryst}}(1-\phi_k). \quad (14)$$

In practice, diffusion coefficients in the solid phases are orders of magnitudes smaller than those in the amorphous phase, so that diffusion inside the crystals is fully negligible over the whole simulation time.

The order parameters obey the classical Allen-Cahn equation,

$$\frac{\partial \Phi_k}{\partial t} = -M_k \left( \frac{\partial \Delta G_V}{\partial \Phi_k} - \nabla \left( \frac{\partial \Delta G_V}{\partial (\nabla \Phi_k)} \right) \right), \qquad (15)$$

Here, $M_k$ is the mobility coefficient for the solid-liquid interface for crystals of material $k$ and will be called "interfacial mobility" in the following. We point out however that in the general case, the crystallization rate obtained in the simulation not only depend on $M_k$, but also on all the thermodynamic and diffusional properties of the system.

The Cahn-Hilliard and the Allen-Cahn equations together ensure that the system progressively relaxes towards its thermodynamic equilibrium, by minimizing its free energy relative to the volume fraction and the order parameter variables.

The second term of the RHS in Equation *(15)* includes a contribution from the orientation mismatch (see Equation *(8)*), so that the evolution of the orientation fields $\theta_k$ has to be calculated as well. In classical OFPF models, the kinetics of crystal orientation is governed by additional Allen-Cahn equations applied to the orientation parameter [65-67] and both order and orientation fields are fully coupled through the last term of Equation *(8)*. This approach has two major drawbacks: first, because of this coupling, the growth rate of an isolated crystal growing within an amorphous environment depends of its orientation, which is non-physical. Second, the interfaces for orientation parameters are much sharper than those for volume fraction and order parameter gradients, inducing a high computational cost. To overcome this drawback, we propose a much simpler and computationally more efficient heuristic procedure for the propagation of the orientation field: when a crystal is growing, the value of the order parameter in the amorphous phase around it increases; when the value of the order parameter on these surrounding nodes exceeds a given threshold value, the nodes are assumed to be crystallizing and are attributed an orientation. The orientation attributed to a given node is simply the one of the (already crystallized) neighboring node with the highest order parameter. As a consequence, the orientation of a given nucleus propagates together with its order parameter field. Within a single crystal, the orientation is thus uniform, so that $|\nabla \theta_k| = 0$ and the orientation mismatch energy term in Equation *(8)* is zero. Moreover, since the orientation is undefined in the amorphous phase, the orientation gradient at a solid-liquid interface is also undefined. In order to ensure that the growth rate of a crystal in a surrounding amorphous phase is independent of its orientation, we simply set $|\nabla \theta_k| = 0$ at the solid-liquid boundaries of the orientation fields. As a consequence, the orientation mismatch contribution is only non-zero at the boundaries between two crystals with different orientations, as desired.

## Expected evolution of BHJ and choice of the simulated systems

In this paper, we investigate the time-dependent morphology of binary mixtures depending on their thermodynamic and kinetic properties. This contributes to the understanding of the stability of OSC bulk-heterojunctions: PAL dry films typically consist of a mixture of one donor and one acceptor material that are often both crystalline with a melting temperature far above the processing and operating temperatures of the cell. This means that there is a thermodynamic driving force pushing the system towards further crystallization which holds until the thermodynamically stable morphology of the PAL is reached (fully or partially crystalline depending whether materials are fully or semi-crystalline).

The morphology obtained at the end of the drying process, or even at the end of the whole fabrication process is in the general case out of equilibrium: the drying process is very fast and crystals do not have sufficient time to grow. At the end of the drying and due to the removal of the solvent, diffusion processes become substantially slower and the structure is kinetically "frozen" in an unstable state. Thermal annealing is a common procedure to provide an opportunity to the PAL for additional evolution [103], for instance in the well-known P3HT-PCBM system [32,104]. Here, crystals can typically grow further, but the annealing time has to be limited to quench once again the structure in the desired state. Otherwise, crystal sizes may increase up to even microns, which is prohibitive for the cell performance. This demonstrates that the thermodynamic equilibrium is still not achieved at the end of the fabrication process. Therefore, during the lifetime of the cell, the system will still be subject to driving forces pushing it towards further crystal growth and purification (if the equilibrium composition of the crystal has not yet been reached). This is in principle similar to what happens during annealing, which can actually be regarded as an accelerated ageing.

The question of the BHJ morphological stability is related to the reliability and robustness of the kinetic quenching. This quenching could be considered as perfect if the rate of phase transformation and diffusion processes would be close to zero. Unless the PAL would be only made of polymers with glass transition temperatures far above the operating temperature, or already fully crystalline, this is in general not the case. For this reason, we expect that during the lifetime of the cell, crystals should grow within a timescale that still has to be identified.

Furthermore, the behavior of the amorphous phase in which the crystals evolve has to be taken into account. From the Flory-Huggins theory, we know that the amorphous phase is stable if the interaction parameter is below a critical limit, so that the crystals grow in a miscible single amorphous phase. Above this limit, the amorphous phase is unstable and spinodal decomposition starts. One might argue that the crystallization process might hinder the spinodal decomposition because the crystal phases are the most stable ones, but this is kinetically very unlikely: spinodal decomposition is a spontaneous process, while the growth process at the crystal surface is thermally activated and requires some ordering at the molecular scale, and hence is expected to be a slower process than LLPS. Actually, it is shown below that the presence of crystals even triggers the LLPS. Thus, in the case of immiscible amorphous phases, the most likely situation is expected to be the following: crystals evolve together within separated amorphous phases with simultaneous crystal growth and amorphous phase coarsening.

The natural question arising from these considerations is the miscibility of real OPV materials, especially in the amorphous phase. According to the Flory-Huggins theory, the interaction parameter is the key to evaluate miscibility. For instance, for a binary system composed of two materials A and B with volume fractions $\varphi_A$ and $\varphi_B$ and a ratio of molar masses $N_A/N_B$, the spinodal curve is defined by $\chi = 0.5(1/N_A\varphi_A + 1/N_B\varphi_B)$. For typical polymer/small molecule systems, this leads to critical values of $\chi$ around 0.6-0.7 at the critical concentration and around 1 for a 1:1 blend. Some efforts have been made to experimentally measure this parameter in well-known OPV systems. [24,39,42,104] However, a proper evaluation is not straightforward and the interaction parameters are not available for many material systems. If they are available, they have often been measured at relatively high temperature. From the reported values that typically range between 0.2 and 1, we can either expect miscibility for some systems such as P3HT with ICBA (Indene-$C_{60}$ Bisadduct), or immiscibility for others such as PCDTBT with PCBM, or PCE11 (PffBT4T-2OD) with PCBM. Nevertheless, there are some discrepancies in the reported values: for the model system P3HT-PCBM, values of $\chi$ between 0.2 and 0.4 at 150°C [104] (thus, expected miscibility at these temperatures) as well as around 0.8 at the same temperature [39] (partial immiscibility expected) have been reported. Moreover, values at room temperature or at the cell operation temperature are not available. Discussing the validity of these values is beyond the scope of this paper and we simply acknowledge these controversies and the lack of data on OPV material miscibility. This is the reason why we cannot claim simulating real systems in this contribution and we limit ourselves to model systems, where both situation of crystal evolution with and without LLPS are investigated.

In order to illustrate some of the various physical properties that might be encountered for different donor/acceptor combinations, we investigate three different kinds of systems: fully amorphous immiscible mixtures (as a reference system), mixtures of two crystalline materials that are miscible in the amorphous phase, and mixtures of two crystalline materials that are immiscible in the amorphous phase. We intentionally consider very simplified model systems, whose material properties might be far from the ones of common experimental donor/acceptor mixtures. This is because the objective of this paper is to demonstrate the ability of our theoretical framework to handle these kinds of problem, and to highlight and focus on the basic physical mechanisms driving the morphology evolution and the interactions between them. Taking into account additional complexity to represent much more realistic systems (asymmetric molar masses or diffusion coefficients, partial crystallinity, anisotropic crystal growth, temperature-dependence of all material properties…) is out of the scope of this paper, but can be handled within our framework in a straightforward way. This will be the topic of future work. Beyond this, performing quantitative simulations for OPV blends however requires careful measurement of the thermodynamic and kinetic properties of the mixture, which is not an easy task as illustrated above. This will also be the topic of future work. Thus we only obtain in this paper a qualitative description of the involved physical processes and don't claim that the conclusions are readily applicable to real systems. The parameters used in the simulations are summarized in **Table 1** to **Table 5** unless specified otherwise in the text.

With a focus on the stability of the PAL, the starting point of the simulation should be the morphology that arises from the fabrication process. Although a global qualitative picture of this morphology has been identified over the years, its precise characteristics remain in general unspecified and are strongly process-dependent: as discussed above, this morphology is an unstable state, partially frozen on the way towards the thermodynamic equilibrium. As a consequence of this ill-defined initial state, we choose the starting point for our simulation to be a homogeneous 1:1 mixture (unless specified differently in the text) with an initial random perturbation of the concentration and (for crystalline systems) 20 randomly distributed nuclei with a radius of 12nm for each material. Further additional nucleation is not taken into account. Thus, in particular for immiscible systems, the state corresponding to the PAL morphology at the end of the process is not the starting point of the simulation, but corresponds to a later moment, typically when the spinodal decomposition has already started. This means that we assume that the PAL morphology corresponds to the structure of our simulated system *shortly* after the start of the simulation, and that the evolution we describe mainly corresponds to the PAL evolution after the drying process.

The initial nuclei are sufficiently large to be stable, i.e. the free energy change upon crystallization dominates the contribution of the surface tension, so that they are expected to grow. The initial radius of 12nm is actually the smallest one for which the nuclei are stable for all the simulated systems.

Note that we perform simulations in the situation where $\Delta G_{V,k}^{cryst} < 0$ and with stable, initial nuclei. As a consequence, crystal tend to grow in general, corresponding to $\partial \Phi_k / \partial t > 0$. Nevertheless, during crystal growth in a multiphase and polycrystalline mixture, interphases with high local curvatures might arise and $\partial \Phi_k / \partial t$ might be locally negative due to the surface tension term, leading to local dissolution of the crystal. However, since crystal surface growth/melting is a thermally activated process and since the free energy of the crystal phase is much smaller than the one of the amorphous phase, the rate of crystal melting is expected to be negligible as compared to the rate of crystal growth. To take this kinetic effect into account, in this work we simply set $M_k=0$ whenever and wherever the RHS of Equation *(15)* becomes negative.

In the simulations presented here, we almost always used very high self-diffusion coefficients of $2 \cdot 10^{-10}$m$^2$/s for all materials. The times needed for microstructure evolution presented in this paper then follow to be around 1 to 10 milliseconds, which is very small as compared to stability-related evolution in OSC. However, the time scale of the simulation is directly proportional to the diffusion coefficients. Expected values of diffusion coefficients in dried films are orders of magnitude lower than those used here. Therefore, for example, diffusion coefficients of about $2 \cdot 10^{-16}$m$^2$/s (this is the order of magnitude for diffusion of PCBM in P3HT at room temperature [105]) would lead to microstructure evolutions over $10^3$ to $10^4$ seconds, diffusion coefficients of about $10^{-20}$m$^2$/s (the order of magnitude for diffusion of C$_{60}$ in PCDTBT around 80°C [106]) to microstructure evolutions of $10^7$ to $10^8$ seconds. However, in reality, donor and acceptor have different, composition-dependent self-diffusion coefficients. Additionally, the crystal growth rate in real systems have not been measured yet. Measuring these growth rates and understanding which diffusion coefficient fixes the time scale for the morphology evolution is currently under investigation and beyond the scope of this paper. Therefore, we do not focus on the time scale for microstructure evolution for now.

The equations are solved in 2 dimensions on a mesh of 512x512 elements. The mesh size has been adapted so that the thinnest encountered interfaces are at least 5 mesh points thick and fixed to 2nm unless specified differently. To ensure numerical stability, we added a numeric contribution to the free energy with the following parameters: $\varphi_{num} = 3.10^{-4}$, $n_{num} = 2$ and $k_{num} = 50$. The equations are numerically solved using an Euler explicit finite difference scheme.

## Amorphous immiscible systems

The case of LLPS phase separation in binary amorphous systems has been studied extensively theoretically [97,98,100,107-109] and numerically [56,84,86]. In this section, we illustrate very briefly this situation in order to provide a reference situation for the case of crystalline immiscible systems described later. We present simulation results for symmetric simple small molecule model systems (see **Table 1**) and for strongly asymmetric polymer-small molecule systems (see **Table 2**). The phase diagrams of these mixtures are shown in **Figure 1**, whereby the simulated systems are marked with a black star. Additionally, for the polymer-small molecule system, the diffusion coefficients are assumed to be composition dependent. Several models have been proposed in the literature for the expression of these coefficients [110]. We propose here to use the equation proposed by Vignes that has both advantages of being a good first order approximation of the well-known dependence of the self-diffusion coefficients of polymers in solution, and of expressing the self-diffusion coefficient depending on the self-diffusion coefficient at infinite dilution $D_{s,i}^{\varphi_k \to 1}$, which are experimentally more accessible:

$$D_{s,i}(\varphi_i) = \prod_{k=1}^{n} \left(D_{s,i}^{\varphi_k \to 1}\right)^{\varphi_k} \tag{16}$$

The diffusion coefficients at infinite dilution are set to be $10^{-13}$, $2 \cdot 10^{-11}$, $2 \cdot 10^{-12}$, $2 \cdot 10^{-10}$ m²/s for the polymer in polymer, polymer in small molecule, small molecule in polymer, small molecule in small molecule, respectively.

| | |
|---|---|
| $T$ | 300 K |
| $v_0$ | $10^{-4}$ m³/mol |
| $N_i$ (all) | 1 |
| $\rho_i$ (all) | 1000 kg/m³ |
| $\chi_{12,ll}$ | 5 |
| $\kappa_i$ (all) | $10^{-10}$ J/m |
| $D_{s,i}^{liq}$ (all) | $2 \cdot 10^{-10}$ m²/s |

**Table 1**: basic parameter set for amorphous immiscible systems with two identical materials

| | |
|---|---|
| $T$ | 300 K |
| $v_0$ | $10^{-4}$ m³/mol |
| $N_i$ | 30 / 1 |
| $\rho_i$ (all) | 1000 kg/m³ |
| $\chi_{12,ll}$ | (See text) |
| $\kappa_i$ (all) | $10^{-10}$ J/m |
| $D_{s,i}^{liq}$ (all) | See text |

**Table 2**: basic parameter set for amorphous immiscible polymer/small molecule system

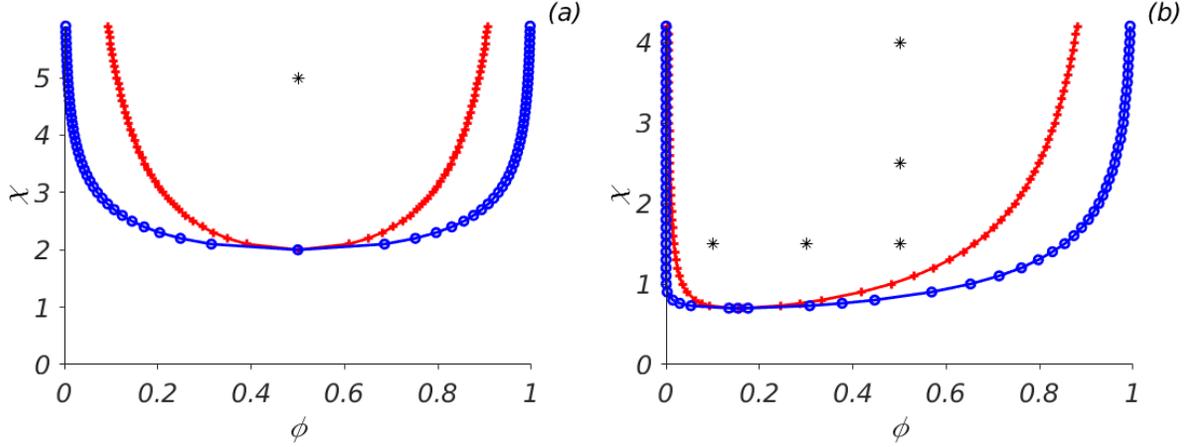

**Figure 1**: phase diagram of the investigated systems, with the spinodal curve (red), the binodal curve (blue) and the simulated systems (black stars) (a) two identical materials (b) polymer-small molecule

| $N_i$ | $\chi_{12,ll}$ | Blend | Volume fraction of material 1 in Phase A | Volume fraction of material 1 in Phase B | Proportion of Phase A |
|---|---|---|---|---|---|
| 1 – 1 | 5 | 1:1 | 0.993 | 0.007 | 50% |
| 30 - 1 | 4 | 1:1 | 0.9925 | $<10^{-4}$ | 50.4% |
| 30 – 1 | 2.5 | 1:1 | 0.961 | $<10^{-4}$ | 52% |
| 30 – 1 | 1.5 | 1:1 | 0.854 | $<10^{-4}$ | 58.5% |
| 30 - 1 | 1.5 | 0.3:0.7 | 0.854 | $<10^{-4}$ | 35.1% |
| 30 - 1 | 1.5 | 0.1:0.9 | 0.854 | $<10^{-4}$ | 11.7% |

**Table 3**: equilibrium properties of the amorphous systems

The thermodynamic parameters and composition of the mixtures are chosen such that they are all unstable and thus demix through spinodal decomposition (**Figure 2a**). The composition of the separated phases as well as their respective proportion in the demixed system can be readily obtained from the binodal curve and the lever rule and are reported in **Table 3**. Remember that if the morphology presented in **Figure 2** for the symmetric system has co-continuous pathways, this is not a general rule: the ability to obtain co-continuous pathways instead of isolated domains of the minority phase into the majority phase depend on the proportion of both phases in the demixed state, and hence from the blend composition, but also on the molecular size and on the interaction parameter. The generated amorphous phases then coarsens (**Figure 2**b). To characterize this coarsening, we obtain the characteristic length scale of the system as follows: the time-dependent structure factor is calculated as the 2D-Fourier transform of the volume fraction field, and integrated over all directions to obtain the probability distribution $p(q,t)$ of $q$-vectors at each time. The characteristic length scale is the inverse of the mean value of $q$ over this distribution, $L(t) = 1/\int qp(q,t)dq$. At late stages, the characteristic domain size $L$-$L_0$ is known to increase in a binary system as $L - L_0 \sim \left(\Lambda_{11} \frac{\partial^2 \Delta G_V^{loc}}{\partial \varphi^2} \sqrt{\kappa \Delta G}(t - t_0)\right)^{1/3}$. Thereby, $\Delta G$ is the change in free energy between the mixed and demixed states, $t_0$ the time when the spinodal decomposition sets on and $L_0$ the characteristic length scale at $t_0$. $t_0$ is evaluated in the following way: the probability distribution of the volume fraction is computed at each time step. As long as the amorphous phase is mixed, the distribution is centered around the blend composition. With the spinodal decomposition setting on, two maxima appear in the distribution corresponding to both separated phases. These maxima quickly reach the expected binodal compositions. $t_0$ is defined as the time from which the composition of the separated phases have reached the binodal compositions within an error of 1%. The result of this procedure is shown in **Figure 3**. Note that this scaling law also holds for the polymer-small molecule system with composition-dependent diffusion coefficients.

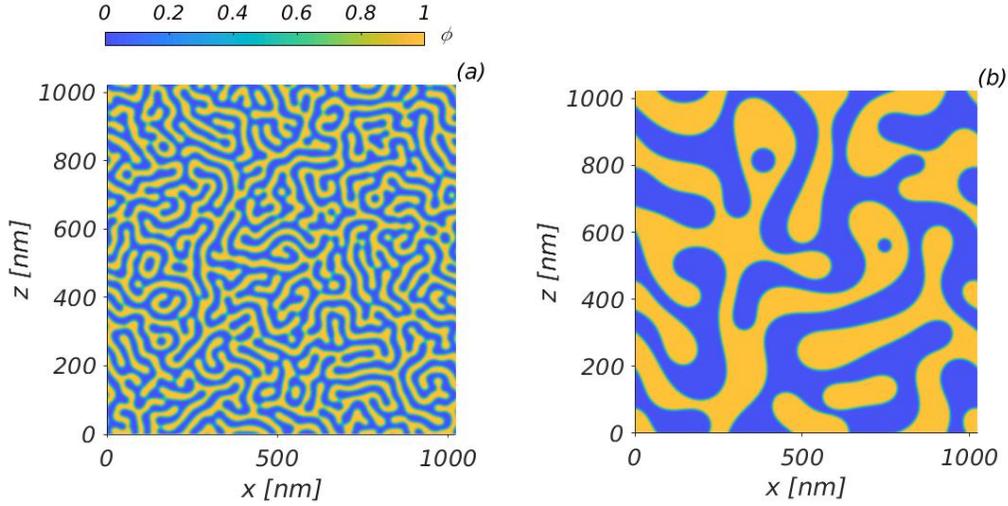

**Figure 2**: Volume fraction field for the 1st material for the amorphous immiscible 1:1 blend with identical materials, (a) t = $5.7 \cdot 10^{-6}$s, (b) $1.6 \cdot 10^{-3}$s

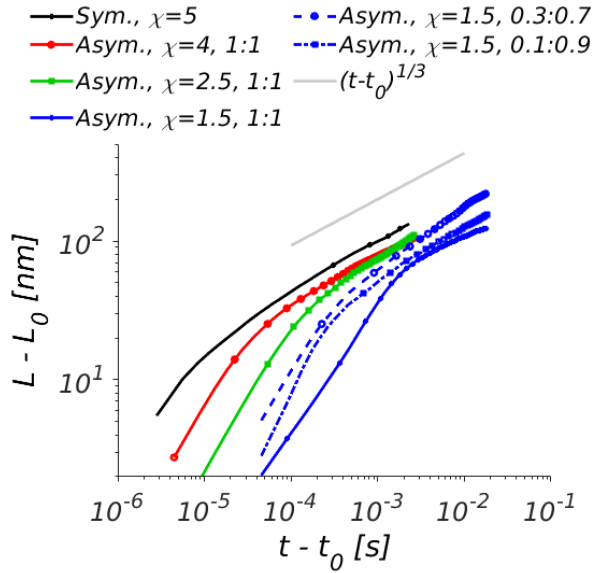

**Figure 3**: Characteristic wavelength computed from the structure factor of the volume fraction field for the amorphous systems; $t_0$ is the time for the onset on spinodal decomposition and $L_0$ the characteristic length scale at $t_0$

# Crystalline miscible systems

In this section, we focus on crystalline miscible systems. Although the thermodynamic properties of the mixture are of primary importance in order to understand the behavior of real OPV systems, we highlight here the importance of purely kinetic properties and the mechanism of crystal development in the evolving structure. The objective is to analyze the interplay between the single crystals and the balance between diffusion rate and crystal growth rate. These findings are of noticeable importance for the time-evolution of classical OPV systems that are composed of three phases with a stable mixed amorphous phase, but also for the understanding of the more complex crystalline immiscible systems (see next section).

The model system we study consists of two miscible materials with identical thermodynamic and kinetic properties. The model parameters are summarized in **Table 4**. The thermodynamic parameters are chosen so that both materials are fully crystalline, and that the solid phases are strongly immiscible. The crystals are assumed to be already highly pure from the beginning of the simulation: the question of the possible composition evolution of crystals that might be still partially mixed with the other active material at the end of the fabrication process is not considered in this paper. This could be investigated with the help of our model in a very straightforward way, however. With the chosen parameter, the amorphous phase is thermodynamically stable, whatever its composition. The thermodynamic stable structure is expected to consist 100% of pure crystals from both materials. In order to analyze the influence of the competition between crystal growth and diffusion processes in the amorphous phase, the parameters are adjusted so that the crystal growth rates are sufficiently high to generate concentration gradients in the amorphous phase. The surface tension parameters $\varepsilon_{g1,k}$ for the orientation gradients are chosen to be sufficiently high in order to avoid crystal coalescence even for very small orientation mismatch.

| $T$ | 300 K | $H_k$ (all) | 12.5 kJ/kg |
|---|---|---|---|
| $v_0$ | $10^{-4}$ m$^3$/mol | $L_k$ (all) | 5 kJ/kg |
| $N_i$ (all) | 1 | $T_{m,k}$ (all) | 600 K |
| $\rho_i$ (all) | 1000 kg/m$^3$ | $\xi_{0,k}$ (all) | 1 |
| $\chi_{12,ll}$ | 1.8 | $\varepsilon_k$ (all) | $5 \cdot 10^{-4}$ (J/m)$^{0.5}$ |
| $\chi_{ij,sl}$ (all) | 6.2 | $\varepsilon_{g1,k}$ (all) | 0.15 (J/m)$^{0.5}$ |
| $\chi_{12,ss}$ | 0 | $\varepsilon_{g2,k}$ (all) | 0 |
| $\kappa_i$ (all) | $10^{-10}$ J/m | $M_k$ (all) | $10^5$ s$^{-1}$ |
| $D_{s,i}^{liq}$ (all) | $2 \cdot 10^{-10}$ m$^2$/s | $k_D$ | 50 |
| | | $\Phi_D$ | 0.5 |

**Table 4**: Basic parameter set for crystalline miscible systems

Before turning to the polycrystalline case, we investigate the growth rate of a single crystal in a mixture of both materials. In that case, the crystal growth rate is constant and it has been established for long that in a pure material, the Allen-Cahn equation results in a growth rate that is strictly proportional to the interfacial mobility. [55,58] Here, we calculate the crystal growth rate depending on composition for different $M$ values, and plot the ratio of the interface velocity to M depending on the volume fraction in **Figure 4**. As expected, the ratio is independent of $M$ in the pure material ($\varphi$=1). For all values of *M*, the crystal does not grow for volume fractions smaller than roughly 25%: below these values, the cost for the increase of the crystal surface due to surface tension is higher than the gain in volume due to the lower free energy in the solid phase, and the nucleus is not stable. Above 25%, the growth rate increases with volume fraction. There is a strong deviation from linearity for low interfacial mobility *M*, that is more pronounced around 50%. This is due to the comparatively stronger $\Phi_1^2 \varphi_1 \varphi_2 \chi_{12,sl}$ solid-liquid interaction term for mixtures with $\varphi$ close to 0.5 in Equation *(4)*. For higher *M* values, the ratio of crystal growth rate to interface mobility becomes lower, except in the pure material. In this regime of high interfacial mobility relative to the diffusion rate, the crystal growth becomes diffusion-limited. This is a consequence of concentration gradients that appear around the crystal, because the diffusion process in the amorphous phase is not fast enough to compensate for the material consumption at the crystal surface. The effective concentration at the surface is therefore reduced, and so the growth rate.

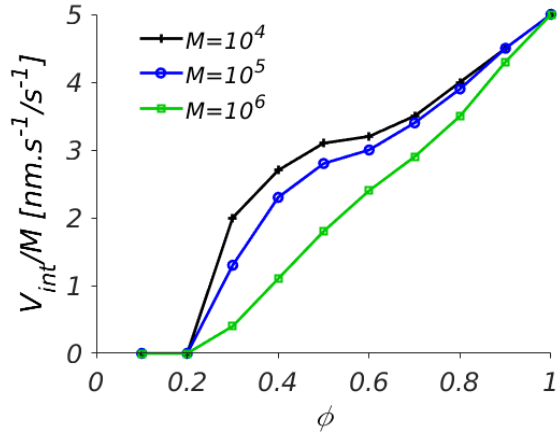

**Figure 4**: Ratio of the growth rate of a single crystal to the interfacial mobility in a binary immiscible system, as a function of volume fraction

We now investigate the case of a polycrystalline 1:1 mixture with the parameters summarized in **Table 4**. The evolution of the volume fraction field for the 1st material, superimposed with the location of the crystallites (yellow for material 1, dark blue for material 2) is shown in **Figure 5**. Spherical crystals develop isotropically until they impinge together (**Figure 5**a and b). The crystal growth is then determined by the available space between the crystals (**Figure 5**c to f). At some point (**Figure 5**d), the solid crystals have almost quenched the topology of the material 1-rich and material 2-rich zones, and the subsequent evolution of the system consists simply of the crystallization of the remaining amorphous domains. The morphology of the final structure (size, topology of the crystals) depends strongly on the location and number of the initial nuclei.

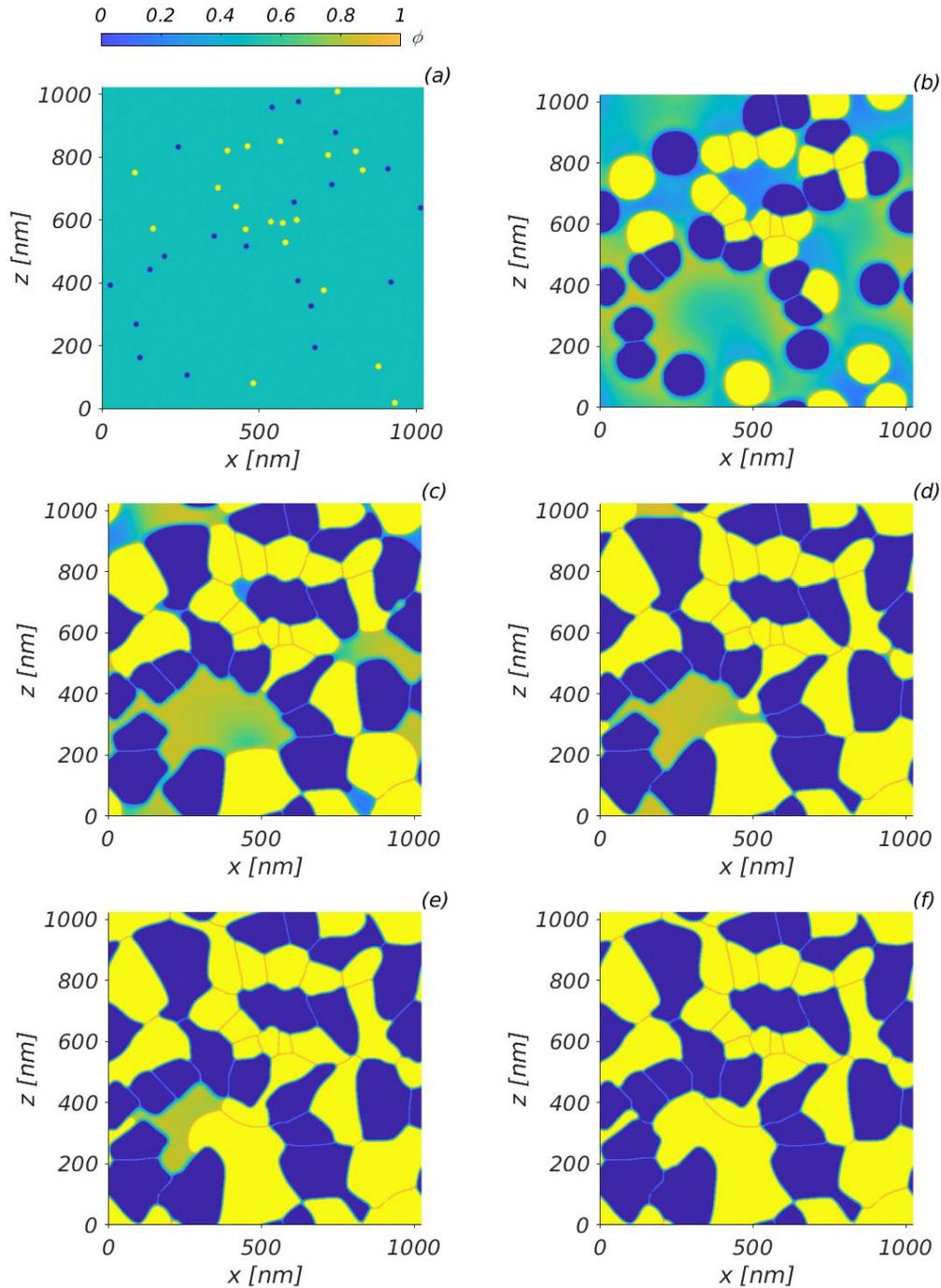

**Figure 5**: Miscible crystalline system, volume fraction field for the 1st material superimposed with the location of the crystallites (yellow for material 1, dark blue for material 2), t = 0s, $2 \cdot 10^{-4}$s, $4 \cdot 10^{-4}$s, $6 \cdot 10^{-4}$s, $8 \cdot 10^{-4}$s, $10^{-3}$s

The crystal growth rate at the beginning of the structure evolution, evaluated through the evolution of the mean equivalent diameter, is exactly the one expected from the single-crystal simulations discussed above (see **Figure 6**). However, the growth rate starts dropping as soon as the crystals impinge. A second contribution to this slow-down is that concentration gradients in amorphous areas surrounded by several crystals are higher than in the single crystal case. This example shows how not only the thermodynamic properties and the relative speed of crystal growth and diffusion processes, but also the overall composition of the mixture and the nucleus density have to be taken into account in order to predict the final film structure.

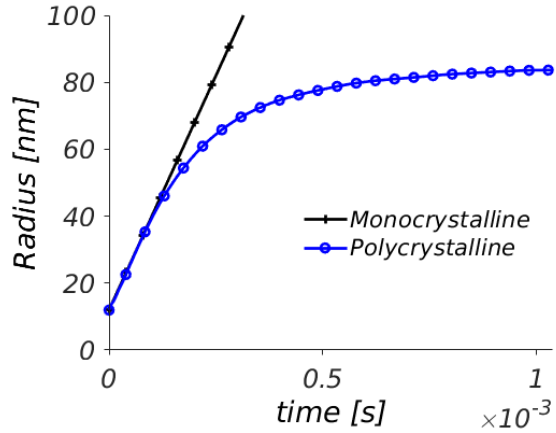

**Figure 6**: comparison of time-dependent crystallite growth rate for a monocrystalline system and a miscible polycrystalline system (mean equivalent crystal radius), 1:1 blend, $M=10^5 s^{-1}$

# Interactions between crystal growth and spinodal decomposition in immiscible systems

In this section, we focus on crystalline immiscible systems. The amorphous phases quickly undergo a spinodal decomposition and the separated liquid phases coarsens while crystals grow at the same time. On top of the previously described processes, the LLPS and the crystal growth influence each other. The objective is to understand the interplay between them and to analyze how the crystals grow in the phase separated fluid.

Once again, the model system we study consists of two immiscible materials with identical thermodynamic and kinetic properties. The model parameters are summarized in **Table 5**. The thermodynamic parameters are chosen so that the liquid phase as well as the solid phases are strongly immiscible, and that both materials are fully crystalline and highly pure in the solid phase. The thermodynamically stable structure is expected to consist exclusively of pure crystals from both materials. In order to illustrate the importance of the kinetics at fixed thermodynamic properties, the crystal growth rate is varied, all other parameters being fixed. This allow us to investigate the influence of the competition between crystal growth and diffusion processes in the amorphous phase on the structure development and final film morphology. Note that to this end, the parameters are adjusted so that the crystal growth rates are relatively high, in the same order of magnitude compared to the amorphous phases coarsening rate.

| $T$ | 300 K | $H_k$ (all) | 5 kJ/kg |
|---|---|---|---|
| $v_0$ | $10^{-4}$ m³/mol | $L_k$ (all) | 2 kJ/kg |
| $N_i$ (all) | 1 | $T_{m,k}$ (all) | 600 K |
| $\rho_i$ (all) | 1000 kg/m³ | $\xi_{0,k}$ (all) | 1 |
| $\chi_{12,ll}$ | 5 | $\varepsilon_k$ (all) | $5\cdot 10^{-4}$ (J/m)$^{0.5}$ |
| $\chi_{ij,sl}$ (all) | 5 | $\varepsilon_{g1,k}$ (all) | 0.15 $(J/m)^{0.5}$ |
| $\chi_{12,ss}$ | 0 | $\varepsilon_{g2,k}$ (all) | 0 |
| $\kappa_i$ (all) | $10^{-10}$ J/m | $M_k$ (all) | See text |
| $D_{s,i}^{liq}$ (all) | $2 \cdot 10^{-10}$ m²/s | $k_D$ | 50 |
| | | $\Phi_D$ | 0.5 |

**Table 5**: Basic parameter set for crystalline immiscible systems

## Crystal development

The evolution of the volume fraction field for the 1st material, superimposed with the location of the crystals (yellow for material 1, dark blue for material 2) is shown in **Figure 7**. The spinodal decomposition onsets at a very short time. Remarkably, the amorphous separated phases are organized in concentric zones around the crystal (**Figure 7**a and b), contrary to the amorphous case (compare with **Figure 2**). This is because the spinodal decomposition is triggered by the concentration gradients around the crystals and not by the Gaussian fluctuations in the amorphous phase. Then, since the growth rate is concentration dependent, the crystals tend to grow in the highly concentrated liquid phases, according to their geometry (**Figure 7**c, d and e). Thus, the crystal growth is guided by the constantly evolving topology of the LLPS. Although the interfacial growth rate is fully isotropic, crystals acquire a highly structured, possibly branched morphology depending on the pathways they find for growth. Since crystals of a given material all grow along the amorphous highly concentrated phases that usually forms percolating pathways for this 1:1 blend, the crystals belonging to a single material also tend to form percolating pathways (**Figure 7**e and f). However, at the same time, growing crystals constantly capture material from the amorphous phases and interfere with their coarsening. They may create separations in the amorphous percolated pathways, and they finally completely quench the coarsening when the crystallinity is sufficiently high to hinder any evolution of the amorphous phases (**Figure 7**e and f). The final state (**Figure 7**f) looks like a classical spinodal LLPS pattern, but this is a solid structure that does not evolve any further. Note again that in all the cases investigated here, the crystal growth rate is high enough so that crystals grow significantly during the coarsening of the amorphous phases. If the crystal growth rate would be very small compared to the rate of amorphous phase coarsening, the spinodal decomposition would start as depicted in **Figure 7**b and c, and then would develop very similar to the fully amorphous case (see **Figure 2** and **Figure 3**).

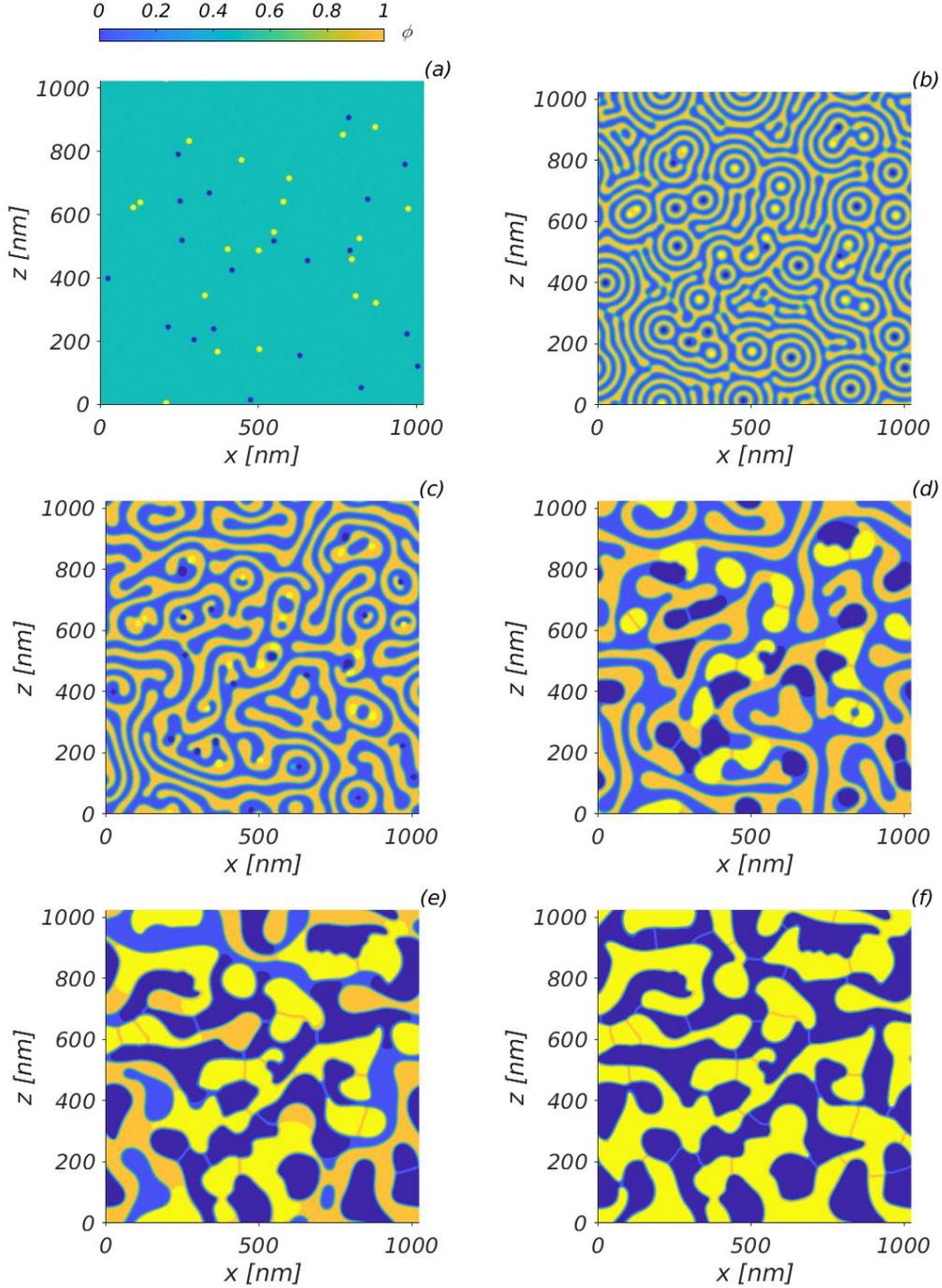

**Figure 7**: immiscible crystalline system with M=$10^5$s$^{-1}$, volume fraction field for the 1$^{st}$ material superimposed with the location of the crystallites (yellow for the 1$^{st}$ material, dark blue for the 2$^{nd}$ material), t = 0s, 5.7·10$^{-6}$s, 4.8·10$^{-5}$s, 1.3·10$^{-4}$s, 2.7·10$^{-4}$s, 2·10$^{-3}$s

The time-dependent growth rate of single crystals shows also remarkable features at the beginning of the structure evolution (see **Figure 8**). We have already outlined that in a pure material system with a single crystal, the interface speed is constant with time and proportional to the interfacial mobility, so that the crystal radius growth linearly with time (full lines in **Figure 8**). In a binary polycrystalline system, three distinct growth phases can be recognized in the evolution of the mean equivalent crystal radius (dashed lines in **Figure 8**), that can be related to particular events in the structure development. At very short times, just after the onset of the spinodal decomposition, crystals absorb all the material around them so that they are quickly surrounded by a circular depleted liquid zone. The growth rate strongly drops, which is the reason for the « plateau » in the crystal radius evolution at small time. This depletion zones disappear with the coarsening of the liquid phases which feed the crystal growth. This effect is less pronounced at low interfacial mobility because crystals grow too slowly to generate a depletion zone around them. In a second phase, single crystals grow isolated from each other in the highly concentrated amorphous phases. The crystal growth rate is very high and close to the growth rate in a pure monocrystalline system, because the separated phase in which the crystal grow are almost pure with the chosen thermodynamic parameters. This holds until the first impingement between crystals. Then, the growth rate progressively decreases with increasing steric hindrance and progressive lack of remaining amorphous material to feed the crystal growth. Note that the mean equivalent radius is not a very good descriptor of the fully crystalline final structure, because the final mean radius is roughly in all cases $\langle R_{end} \rangle = \sqrt{S/2\pi n}$, whereby $S$ is the system size, $n$ the number of crystals and the factor 2 appears because we consider 1:1 blends. The dependence of the final structure on the crystallization rate will be discussed in detail in the next section.

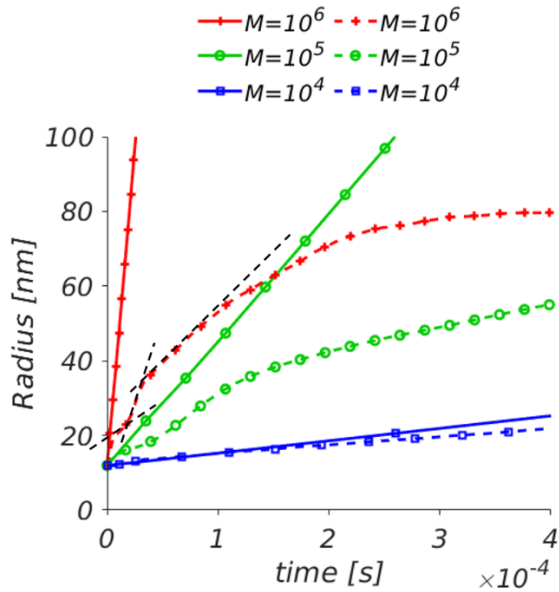

**Figure 8**: Comparison of time-dependent growth rate for a single crystal in a pure material (full lines) and an immiscible polycrystalline system (mean equivalent crystal radius, dotted lines) for a 1:1 blend, at early times. The black dashed lines are guide to the eye for the three growth phases.

The overall crystallization rate (defined as the derivative of the total crystalline volume with respect to time) is shown in **Figure 9**. Remarkably, the higher the interfacial mobility, the more irregular the overall growth rate. This is because for high interfacial mobilities, crystals are more likely to consume the material in highly concentrated amorphous phases around them, creating depletion zones and a sudden drop of their growth rate, until a new highly concentrated phase form around them and the growth can be very fast once again. At low interfacial mobility, the growth rate becomes much smoother, close to the situation of the miscible system investigated in the previous section (black curve).

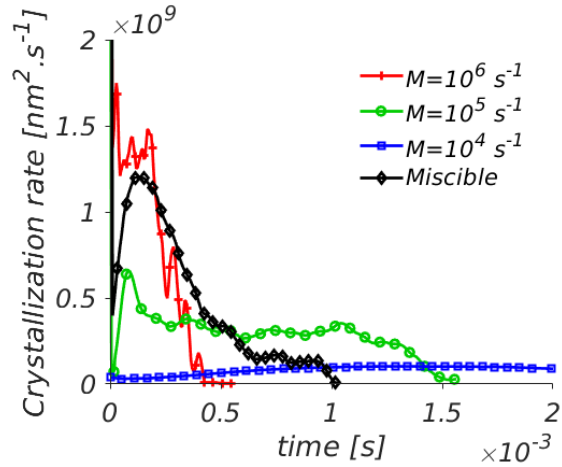

**Figure 9**: Comparison of time-dependent overall crystallization rate for immiscible systems with different interfacial mobilities, and for a miscible system

## Influence of the crystallization rate on the final structure

In this section, we study the impact of the interfacial mobility on the final, stable film structure. The order parameter of the final state for interfacial mobilities ranging from $M=3 \cdot 10^4 s^{-1}$ to $M=10^6 s^{-1}$ are shown in **Figure 10**. Here, in order to visualize the order parameter of both materials on the same figure, the order parameter field for material 2 has been set to vary from 0 (amorphous) to -1 (crystalline) and both order parameter fields are added together. The higher the interfacial mobility, the more structured the final single crystals and the polycrystalline morphology: with higher interface mobility, crystals develop faster and hence following a still very finely interpenetrated amorphous phase separated mixture. In parallel, the final structure is finer because the crystal kinetically quench the system before the liquid phases have time to coarsen.

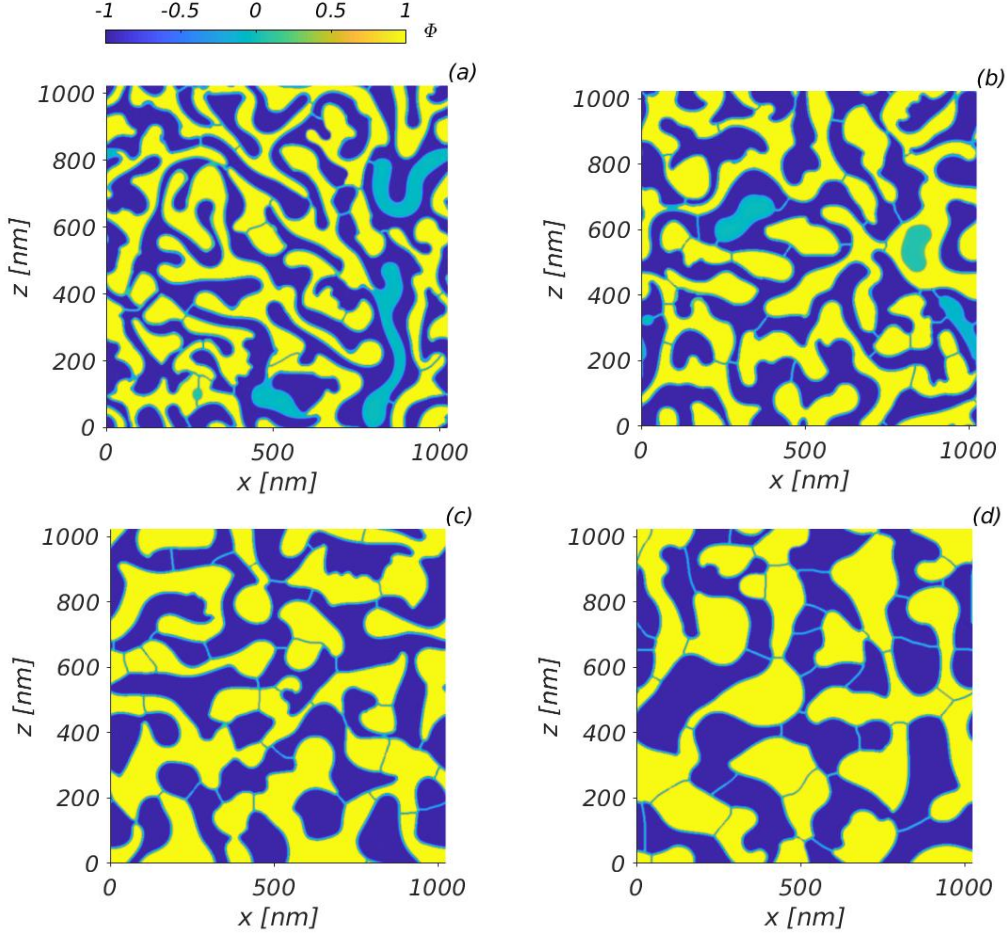

**Figure 10**: Order parameter field of the final structure, ranging from -1 (dark blue, crystal of the 2$^{nd}$ material) to 0 (amorphous) and 1 (dark blue, crystal of the 1$^{st}$ material)
for (a) M = $10^6$ s$^{-1}$, (b) M = $3·10^5$ s$^{-1}$, (c) M = $10^5$ s$^{-1}$, (d) M = $3·10^4$ s$^{-1}$

This can be observed by plotting the time-dependent characteristic wavelength of the system computed from the structure factor of the volume fraction field (**Figure 11**). Whereas the characteristic size of the domains $(L^3-L_0^3)^{1/3}$ increases as $t^{1/3}$ in an amorphous system, it suddenly reaches an asymptotical value in crystalline systems, when the presence of the crystals quenches any further coarsening of the amorphous phases. This quench occurs at shorter times, and therefore generates shorter characteristic sizes if the interfacial mobility is higher. The highest characteristic length is reached in the miscible system investigated in the previous section, for which the crystal growth is not influenced by the LLPS.

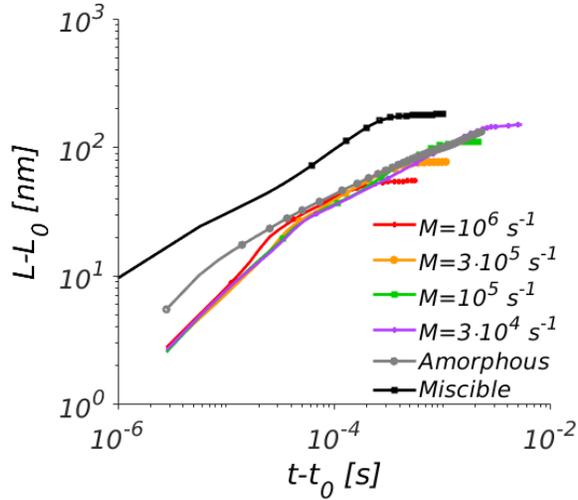

**Figure 11**: Characteristic wavelength computed from the structure factor of the volume fraction field for an amorphous symmetric system, a crystalline miscible systems and crystalline immiscible systems with different interface mobility; $t_0$ is the time for the onset of spinodal decomposition and $L_0$ the characteristic length scale at $t_0$

Furthermore, it can be observed that for high interfacial mobility (**Figure 10**a and b), some areas remain amorphous, although it would be expected from thermodynamic parameters that the whole system would crystallize. In fact, these areas can be considered as amorphous "defects" in the crystalline structure that are kinetically generated. They can form for instance if an amorphous volume happens to be fully surrounded by a crystal of the other material, thus being inaccessible for its own crystallization to take place (unless new nuclei might form, which has not been taken into account in this work). This is the case for the simulation presented in **Figure 10**b. They also can form even if a crystal could in principle topologically reach the remaining amorphous zone, but should go through a bottleneck (**Figure 10**a). This bottleneck generates a surface tension that is too high for the crystal to grow further. Even if the morphology is not in thermodynamic equilibrium, they are long-living metastable structures, the defects being kinetically quenched as long as the solid crystals can be considered as fixed. It should be emphasized that the appearance of the defects depends on the location of the initial nuclei and is not systematic. Nevertheless, the probability of existence of such defects increases with interfacial mobility.

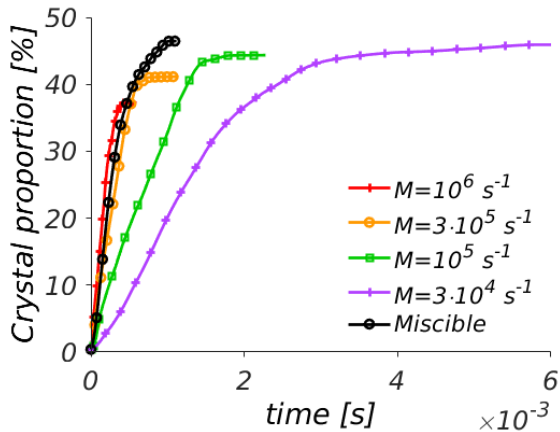

**Figure 12**: Crystalline volume of the 1st material related to the whole volume for a crystalline miscible systems and crystalline immiscible systems with different interfacial mobility

**Figure 12** shows the crystallinity of the 1st material (the evolution for the 2nd material is similar), defined by the total crystalline volume of this material normalized by the total system volume. In such a 1:1 blend, the final crystallinity is expected to be a little bit less than 50%, taking into account the volume of the not fully crystalline interfaces. The crystallinity reaches its final value after the characteristic wavelength has stabilized, meaning that the phase morphology is fully quenched before the crystals are completely grown. The presence of the defects can be recognized in the lower final crystallinity values at higher interfacial mobilities. At last, **Figure 13** shows the time needed to reach the final morphology, which is a good indicator of the stability performance of the mixture, depending on the interfacial mobility. It is found that it follows a power law very close to $M^{-3/4}$, the best fit being $M^{-0.735}$ with an $R^2$ value of 0.99. Further investigation and simulations on more systems with different parameters are needed to investigate the generality of this finding, and to understand how the time to equilibrium depends on the other properties of the system.

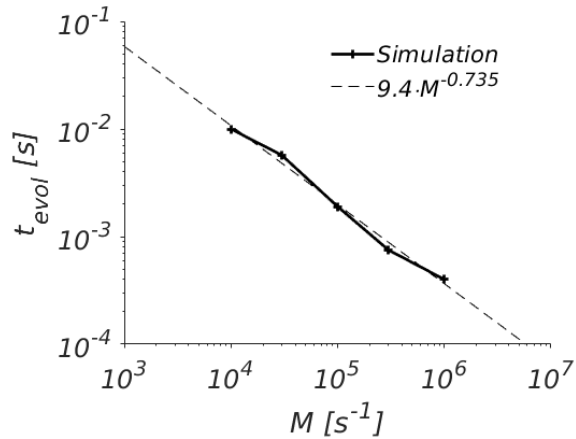

**Figure 13**: Time needed to reach the final morphology, depending on the interfacial mobility

## Conclusions and perspectives

The kinetic evolution of immiscible crystalline mixtures has been rarely investigated theoretically in the literature. However, it is of primary importance in order to understand the evolution of the photoactive layers of organic solar cells during post-processing and operation, so that the stability of these bulk-heterojunctions can be improved. In this paper, we presented a new phase-field simulation framework for the investigation of such systems. The free energy functional contains the description of the mixing term through a Flory-Huggins contribution, as well as the free energy change upon crystallization for each crystalline material and gradient terms generating surface tension effects. This allows to handle liquid-solid phase changes and liquid-liquid demixing at the same time. The volume fraction fields and the order parameter fields evolve towards the thermodynamic equilibrium via the Cahn-Hilliard and the Allen-Cahn equation, respectively. Additionally, the orientation of each crystal is taken into account, generating crystal impingement through a misorientation energetic contribution in the free energy. The proposed simulation code is three-dimensional, although only 2D simulations have been presented in this paper, and it can handle any number of materials. We used this model to investigate the evolution of crystalline miscible and immiscible binary systems, in order to better understand the stability of bulk-heterojunction photoactive layers.

In systems where both donor and acceptor materials are crystalline but miscible in the amorphous phase, the mechanisms of polycrystalline growth have been highlighted. Compared to the growth rate of a single crystal in a pure material, the growth rate in the binary system drops with the concentration. It can also become diffusion-limited if the interfacial growth rate is high enough because depleted zones then surround the crystal, limiting the amount of material available for growth. Finally, impingement between crystals limits the growth rate and are responsible for highly anisotropic single crystals, although the growth rate is fully isotropic.

In immiscible binary systems, the liquid-liquid demixing and the crystallization processes strongly interact together. Starting from a homogenous amorphous phase, the spinodal decomposition is induced by the crystal nuclei and the amorphous separated phases organize around them at short times. Material consumption through crystal growth perturbs the coarsening of the amorphous phases. With increasing crystallinity, this coarsening becomes progressively quenched by the presence of the solid crystals. Conversely, the crystal growth is induced by the spinodal decomposition: it is driven by the local concentration, so that the crystal growth rate and direction are given by the topology of the separated amorphous phases surrounding the crystal. In such systems, the increase of the overall crystallinity is highly irregular, depending on whether a highly concentrated amorphous zone is available around the crystal at each time. Furthermore, the effect of the interfacial growth rate on the final structure is remarkable. With a faster crystallization, the single crystals are more structured and form percolating pathways for each material with smaller lateral dimensions. Moreover, the higher the growth rate, the more amorphous areas can be found in the final structure, which can be considered as crystallinity defects. These findings might be crucial for the efficiency of solar cells, whereby the dimensions of the phases should remain in the order of some tenths of nanometers to ensure exciton dissociation, and for which pathways to the respective electrodes should be available for both electrons and holes.

In a future work, this model will be improved to take into account further nucleation of the crystals as well as crystal growth anisotropy and partial crystallinity which are typically encountered for polymeric donor materials. Complementary to this work where we dealt with simple model systems, it will be used to investigate realistic OPV donor/acceptor polymer/small molecule systems. Thereby, the first challenge is to obtain reasonable input parameters from experimental measurements. In particular, the phase diagrams of the systems should be identified [23,24,26,42], focusing on the question whether liquid-liquid spinodal decomposition is expected. Nevertheless, this work also shows that the knowledge of the kinetic parameters (crystal growth rate, diffusion coefficients but also nucleation rate) is very important to quantitatively investigate the morphology evolution. Once the input parameters are available, a careful validation of the model has to be performed by extensive comparison of the simulated structures with experimental measurements. The objective of this approach will be to assess the relevance of our phase-field framework for investigating OPV bulk heterojunctions. If the validation is successful, it could be used to draw conclusions on the stability behavior of real OPV systems.

# Acknowledgements

The authors acknowledge financial support by the Deutsche Forschungsgemeinschaft (DFG) within the Cluster of Excellence "Engineering of Advanced Materials" (project EXC 315) (Bridge Funding), and gratefully thank Dr. Olga Wodo for fruitful discussions.

# Literature

[1] Y. Liu, J. Zhao, Z. Li, C. Mu, W. Ma, H. Hu, K. Jiang, H. Lin, H. Ade, H. Yan, *Nature Communications* **2014**, *5*, DOI 10.1038/ncomms6293.
[2] Z. Xiao, X. Jia, L. Ding, *Science Bulletin* **2017**, *62*, 1562–1564.
[3] W. Zhao, S. Li, H. Yao, S. Zhang, Y. Zhang, B. Yang, J. Hou, *Journal of the American Chemical Society* **2017**, *139*, 7148–7151.
[4] J. Yuan, Y. Zhang, L. Zhou, G. Zhang, H.-L. Yip, T.-K. Lau, X. Lu, C. Zhu, H. Peng, P. A. Johnson, et al., *Joule* **2019**, *3*, 1140–1151.
[5] Y. Cui, H. Yao, J. Zhang, T. Zhang, Y. Wang, L. Hong, K. Xian, B. Xu, S. Zhang, J. Peng, et al., *Nat Commun* **2019**, *10*, 2515.
[6] L. Hong, H. Yao, Z. Wu, Y. Cui, T. Zhang, Y. Xu, R. Yu, Q. Liao, B. Gao, K. Xian, et al., *Adv. Mater.* **2019**, *31*, 1903441.
[7] "Champion Photovoltaic Module Efficiency Chart | Photovoltaic Research | NREL," can be found under https://www.nrel.gov/pv/module-efficiency.html, **n.d.**
[8] S. A. Gevorgyan, I. M. Heckler, E. Bundgaard, M. Corazza, M. Hösel, R. R. Søndergaard, G. A. dos Reis Benatto, M. Jørgensen, F. C. Krebs, *J. Phys. D: Appl. Phys.* **2017**, *50*, 103001.
[9] X. Du, T. Heumueller, W. Gruber, A. Classen, T. Unruh, N. Li, C. J. Brabec, *Joule* **2019**, *3*, 215–226.
[10] Q. Wang, Y. Xie, F. Soltani-Kordshuli, M. Eslamian, *Renewable and Sustainable Energy Reviews* **2016**, *56*, 347–361.
[11] W. R. Mateker, M. D. McGehee, *Adv. Mater.* **2017**, *29*, 1603940.
[12] A. Uddin, M. Upama, H. Yi, L. Duan, *Coatings* **2019**, *9*, 65.
[13] G. Griffini, S. Turri, *J. Appl. Polym. Sci.* **2016**, *133*, n/a-n/a.
[14] A. Distler, T. Sauermann, H.-J. Egelhaaf, S. Rodman, D. Waller, K.-S. Cheon, M. Lee, D. M. Guldi, *Advanced Energy Materials* **2014**, *4*, 1300693.
[15] C. Müller, T. A. M. Ferenczi, M. Campoy-Quiles, J. M. Frost, D. D. C. Bradley, P. Smith, N. Stingelin-Stutzmann, J. Nelson, *Advanced Materials* **2008**, *20*, 3510–3515.
[16] N. D. Treat, A. Varotto, C. J. Takacs, N. Batara, M. Al-Hashimi, M. J. Heeney, A. J. Heeger, F. Wudl, C. J. Hawker, M. L. Chabinyc, *Journal of the American Chemical Society* **2012**, *134*, 15869–15879.
[17] M. Scarongella, J. De Jonghe-Risse, E. Buchaca-Domingo, M. Causa', Z. Fei, M. Heeney, J.-E. Moser, N. Stingelin, N. Banerji, *Journal of the American Chemical Society* **2015**, *137*, 2908–2918.
[18] X. Du, X. Jiao, S. Rechberger, J. D. Perea, M. Meyer, N. Kazerouni, E. Spiecker, H. Ade, C. J. Brabec, R. H. Fink, et al., *Macromolecules* **2017**, *50*, 2415–2423.
[19] J. A. Bartelt, Z. M. Beiley, E. T. Hoke, W. R. Mateker, J. D. Douglas, B. A. Collins, J. R. Tumbleston, K. R. Graham, A. Amassian, H. Ade, et al., *Advanced Energy Materials* **2013**, *3*, 364–374.
[20] O. Wodo, S. Tirthapura, S. Chaudhary, B. Ganapathysubramanian, *Organic Electronics* **2012**, *13*, 1105–1113.
[21] O. Wodo, J. D. Roehling, A. J. Moulé, B. Ganapathysubramanian, *Energy Environ. Sci.* **2013**, *6*, 3060.
[22] O. Wodo, J. Zola, B. S. Sarath Pokuri, P. Du, B. Ganapathysubramanian, *Materials Discovery* **2015**, *1*, 21–28.
[23] L. Ye, B. A. Collins, X. Jiao, J. Zhao, H. Yan, H. Ade, *Advanced Energy Materials* **2018**, 1703058.
[24] L. Ye, H. Hu, M. Ghasemi, T. Wang, B. A. Collins, J.-H. Kim, K. Jiang, J. H. Carpenter, H. Li, Z. Li, et al., *Nature Materials* **2018**, *17*, 253–260.

[25] B. S. S. Pokuri, J. Sit, O. Wodo, D. Baran, T. Ameri, C. J. Brabec, A. J. Moule, B. Ganapathysubramanian, *Adv. Energy Mater.* **2017**, *7*, 1701269.
[26] N. Li, F. Machui, D. Waller, M. Koppe, C. J. Brabec, *Solar Energy Materials and Solar Cells* **2011**, *95*, 3465–3471.
[27] N. S. Güldal, T. Kassar, M. Berlinghof, T. Ameri, A. Osvet, R. Pacios, G. Li Destri, T. Unruh, C. J. Brabec, *Journal of Materials Chemistry C* **2016**, *4*, 2178–2186.
[28] B. Schmidt-Hansberg, M. Sanyal, M. F. G. Klein, M. Pfaff, N. Schnabel, S. Jaiser, A. Vorobiev, E. Müller, A. Colsmann, P. Scharfer, et al., *ACS Nano* **2011**, *5*, 8579–8590.
[29] B. Schmidt-Hansberg, M. F. G. Klein, M. Sanyal, F. Buss, G. Q. G. de Medeiros, C. Munuera, A. Vorobiev, A. Colsmann, P. Scharfer, U. Lemmer, et al., *Macromolecules* **2012**, *45*, 7948–7955.
[30] R. S. Gebhardt, P. Du, O. Wodo, B. Ganapathysubramanian, *Computational Materials Science* **2017**, *129*, 220–225.
[31] N. Li, J. D. Perea, T. Kassar, M. Richter, T. Heumueller, G. J. Matt, Y. Hou, N. S. Güldal, H. Chen, S. Chen, et al., *Nature Communications* **2017**, *8*, 14541.
[32] S. A. Dowland, M. Salvador, J. D. Perea, N. Gasparini, S. Langner, S. Rajoelson, H. H. Ramanitra, B. D. Lindner, A. Osvet, C. J. Brabec, et al., *ACS Applied Materials & Interfaces* **2017**, *9*, 10971–10982.
[33] C. Zhang, A. Mumyatov, S. Langner, J. D. Perea, T. Kassar, J. Min, L. Ke, H. Chen, K. L. Gerasimov, D. V. Anokhin, et al., *Advanced Energy Materials* **2017**, *7*, 1601204.
[34] C. Zhang, T. Heumueller, S. Leon, W. Gruber, K. Burlafinger, X. Tang, J. D. Perea, I. Wabra, A. Hirsch, T. Unruh, et al., *Energy Environ. Sci.* **2019**, *12*, 1078–1087.
[35] Y. Zhu, A. Gadisa, Z. Peng, M. Ghasemi, L. Ye, Z. Xu, S. Zhao, H. Ade, *Adv. Energy Mater.* **2019**, *9*, 1900376.
[36] N. Gasparini, M. Salvador, S. Strohm, T. Heumueller, I. Levchuk, A. Wadsworth, J. H. Bannock, J. C. de Mello, H.-J. Egelhaaf, D. Baran, et al., *Adv. Energy Mater.* **2017**, *7*, 1700770.
[37] L. Ye, S. Li, X. Liu, S. Zhang, M. Ghasemi, Y. Xiong, J. Hou, H. Ade, *Joule* **2019**, *3*, 443–458.
[38] J. D. Perea, S. Langner, M. Salvador, J. Kontos, G. Jarvas, F. Winkler, F. Machui, A. Görling, A. Dallos, T. Ameri, et al., *The Journal of Physical Chemistry B* **2016**, *120*, 4431–4438.
[39] J. D. Perea, S. Langner, M. Salvador, B. Sanchez-Lengeling, N. Li, C. Zhang, G. Jarvas, J. Kontos, A. Dallos, A. Aspuru-Guzik, et al., *The Journal of Physical Chemistry C* **2017**, *121*, 18153–18161.
[40] K. Zhou, J. Liu, M. Li, X. Yu, R. Xing, Y. Han, *J. Phys. Chem. C* **2015**, *119*, 1729–1736.
[41] Z. Peng, X. Jiao, L. Ye, S. Li, J. J. Rech, W. You, J. Hou, H. Ade, *Chem. Mater.* **2018**, *30*, 3943–3951.
[42] S. Nilsson, A. Bernasik, A. Budkowski, E. Moons, *Macromolecules* **2007**, *40*, 8291–8301.
[43] J.-H. Kim, A. Gadisa, C. Schaefer, H. Yao, B. R. Gautam, N. Balar, M. Ghasemi, I. Constantinou, F. So, B. T. O'Connor, et al., *Journal of Materials Chemistry A* **2017**, *5*, 13176–13188.
[44] M. Kim, J. Lee, S. B. Jo, D. H. Sin, H. Ko, H. Lee, S. G. Lee, K. Cho, *J. Mater. Chem. A* **2016**, *4*, 15522–15535.
[45] F. Liu, S. Ferdous, E. Schaible, A. Hexemer, M. Church, X. Ding, C. Wang, T. P. Russell, *Advanced Materials* **2015**, *27*, 886–891.
[46] C.-K. Lee, O. Wodo, B. Ganapathysubramanian, C.-W. Pao, *ACS Applied Materials & Interfaces* **2014**, *6*, 20612–20624.

[47] C.-K. Lee, C.-W. Pao, *ACS Applied Materials & Interfaces* **2016**, *8*, 20691–20700.
[48] H. Yoo, K. C. Kim, S. S. Jang, *Computational Materials Science* **2017**, *126*, 299–307.
[49] C. Du, Y. Ji, J. Xue, T. Hou, J. Tang, S.-T. Lee, Y. Li, *Scientific Reports* **2015**, *5*, DOI 10.1038/srep16854.
[50] X. Xu, Y. Ji, C. Du, T. Hou, Y. Li, *RSC Advances* **2015**, *5*, 70939–70948.
[51] I. V. Neratova, A. S. Pavlov, P. G. Khalatur, *Polymer Science Series A* **2010**, *52*, 959–969.
[52] K. Ch. Daoulas, M. Müller, J. J. de Pablo, P. F. Nealey, G. D. Smith, *Soft Matter* **2006**, *2*, 573–583.
[53] D. Kipp, V. Ganesan, *The Journal of Physical Chemistry B* **2014**, *118*, 4425–4441.
[54] N. Moelans, B. Blanpain, P. Wollants, *Calphad* **2008**, *32*, 268–294.
[55] N. Provatas, K. Elder, **n.d.**, 264.
[56] E. B. Nauman, D. Q. He, *Chemical Engineering Science* **2001**, 20.
[57] S. B. Biner, *Programming Phase-Field Modeling*, Springer International Publishing, Cham, **2017**.
[58] T. Takaki, *ISIJ International* **2014**, *54*, 437–444.
[59] H. Xu, R. Matkar, T. Kyu, *Physical Review E* **2005**, *72*, DOI 10.1103/PhysRevE.72.011804.
[60] J. P. Simmons, Y. Wen, C. Shen, Y. Z. Wang, *Materials Science and Engineering: A* **2004**, *365*, 136–143.
[61] I. Loginova, J. Ågren, G. Amberg, *Acta Materialia* **2004**, *52*, 4055–4063.
[62] J. A. Warren, T. Pusztai, L. Környei, L. Gránásy, *Physical Review B* **2009**, *79*, DOI 10.1103/PhysRevB.79.014204.
[63] G. Tegze, T. Pusztai, G. Tóth, L. Gránásy, A. Svandal, T. Buanes, T. Kuznetsova, B. Kvamme, *The Journal of Chemical Physics* **2006**, *124*, 234710.
[64] L. Gránásy, T. Börzsönyi, T. Pusztai, *Phys. Rev. Lett.* **2002**, *88*, 206105.
[65] J. A. Warren, R. Kobayashi, A. E. Lobkovsky, W. Craig Carter, *Acta Materialia* **2003**, *51*, 6035–6058.
[66] L. Gránásy, T. Pusztai, J. A. Warren, *Journal of Physics: Condensed Matter* **2004**, *16*, R1205–R1235.
[67] L. Gránásy, L. Rátkai, A. Szállás, B. Korbuly, G. I. Tóth, L. Környei, T. Pusztai, *Metallurgical and Materials Transactions A* **2014**, *45*, 1694–1719.
[68] B. Korbuly, T. Pusztai, G. I. Tóth, H. Henry, M. Plapp, L. Gránásy, *Journal of Crystal Growth* **2017**, *457*, 32–37.
[69] A. A. Alfarraj, E. B. Nauman, *Macromolecular Theory and Simulations* **2007**, *16*, 627–631.
[70] R. Saxena, G. T. Caneba, *Polym. Eng. Sci.* **2002**, *42*, 1019–1031.
[71] Y. Shang, D. Kazmer, M. Wei, J. Mead, C. Barry, *The Journal of Chemical Physics* **2008**, *128*, 224909.
[72] Y. Shang, L. Fang, M. Wei, C. Barry, J. Mead, D. Kazmer, *Polymer* **2011**, *52*, 1447–1457.
[73] C. Huang, M. O. de la Cruz, B. W. Swift, *Macromolecules* **1995**, *28*, 7996–8005.
[74] C. Huang, M. Olvera de la Cruz, *Macromolecules* **1994**, *27*, 4231–4241.
[75] M. Vonka, J. Kosek, *Chemical Engineering Journal* **2012**, *207–208*, 895–905.
[76] M. R. Cervellere, Y. Tang, X. Qian, D. M. Ford, P. C. Millett, *Journal of Membrane Science* **2019**, *577*, 266–273.
[77] O. Wodo, B. Ganapathysubramanian, *Journal of Computational Physics* **2011**, *230*, 6037–6060.
[78] O. Wodo, B. Ganapathysubramanian, *Computational Materials Science* **2012**, *55*, 113–126.
[79] S. Kouijzer, J. J. Michels, M. van den Berg, V. S. Gevaerts, M. Turbiez, M. M. Wienk, R. A. J. Janssen, *Journal of the American Chemical Society* **2013**, *135*, 12057–12067.
[80] J. J. Michels, E. Moons, *Macromolecules* **2013**, *46*, 8693–8701.

[81] C. Schaefer, P. van der Schoot, J. J. Michels, *Physical Review E* **2015**, *91*, DOI 10.1103/PhysRevE.91.022602.
[82] C. Schaefer, *Theory of Nanostructuring in Solvent-Deposited Thin Polymer Films*, Technische Universiteit Eindhoven, **2016**.
[83] V. Negi, O. Wodo, J. J. van Franeker, R. A. J. Janssen, P. A. Bobbert, *ACS Appl. Energy Mater.* **2018**, *1*, 725–735.
[84] B. Ray, P. R. Nair, M. A. Alam, *Solar Energy Materials and Solar Cells* **2011**, *95*, 3287–3294.
[85] B. Ray, M. A. Alam, *Solar Energy Materials and Solar Cells* **2012**, *99*, 204–212.
[86] M. A. Alam, B. Ray, M. R. Khan, S. Dongaonkar, *Journal of Materials Research* **2013**, *28*, 541–557.
[87] D. Zhou, A.-C. Shi, P. Zhang, *The Journal of Chemical Physics* **2008**, *129*, 154901.
[88] P. Rathi, T. Kyu, *Physical Review E* **2009**, *79*, DOI 10.1103/PhysRevE.79.031802.
[89] D. M. Saylor, C.-S. Kim, D. V. Patwardhan, J. A. Warren, *Acta Biomaterialia* **2007**, *3*, 851–864.
[90] C.-S. Kim, D. M. Saylor, M. K. McDermott, D. V. Patwardhan, J. A. Warren, *Journal of Biomedical Materials Research Part B: Applied Biomaterials* **2009**, *90B*, 688–699.
[91] D. M. Saylor, J. E. Guyer, D. Wheeler, J. A. Warren, *Acta Biomaterialia* **2011**, *7*, 604–613.
[92] M. K. Mitra, M. Muthukumar, *The Journal of Chemical Physics* **2010**, *132*, 184908.
[93] M. Nagai, J. Huang, T. Zhou, W. Huang, *Journal of Polymer Science Part B: Polymer Physics* **2017**, *55*, 1273–1277.
[94] P. J. Flory, *Principles of Polymer Chemistry*, Cornell University Press, **1953**.
[95] R. A. Matkar, T. Kyu, *The Journal of Physical Chemistry B* **2006**, *110*, 12728–12732.
[96] R. A. Matkar, T. Kyu, *The Journal of Physical Chemistry B* **2006**, *110*, 16059–16065.
[97] J. W. Cahn, J. E. Hilliard, *The Journal of Chemical Physics* **1958**, *28*, 258–267.
[98] J. W. Cahn, *Acta Metallurgica* **1961**, *9*, 795–801.
[99] G. A. Buxton, N. Clarke, *Europhysics Letters (EPL)* **2007**, *78*, 56006.
[100] P. G. de Gennes, *The Journal of Chemical Physics* **1980**, *72*, 4756–4763.
[101] E. J. Kramer, P. Green, C. J. Palmstrøm, *Polymer* **1984**, *25*, 473–480.
[102] O. J. J. Ronsin, J. Harting, *Macromolecules* **2019**, *52*, 6035–6044.
[103] F. Zhao, C. Wang, X. Zhan, *Advanced Energy Materials* **2018**, 1703147.
[104] F. Liu, D. Chen, C. Wang, K. Luo, W. Gu, A. L. Briseno, J. W. P. Hsu, T. P. Russell, *ACS Appl. Mater. Interfaces* **2014**, 12.
[105] N. D. Treat, T. E. Mates, C. J. Hawker, E. J. Kramer, M. L. Chabinyc, *Macromolecules* **2013**, *46*, 1002–1007.
[106] C. Saller, F.-J. Kahle, T. Müller, T. Hahn, S. Tscheuschner, D. Priadko, P. Strohriegl, H. Bässler, A. Köhler, *ACS Appl. Mater. Interfaces* **2018**, *10*, 21499–21509.
[107] P. Pincus, *The Journal of Chemical Physics* **1981**, *75*, 1996–2000.
[108] J. H. Yao, K. R. Elder, H. Guo, M. Grant, *Phys. Rev. B* **1993**, *47*, 14110–14125.
[109] A. Baldan, **n.d.**, 32.
[110] C. Peters, L. Wolff, T. J. H. Vlugt, A. Bardow, in *Experimental Thermodynamics Volume X* (Eds.: D. Bedeaux, S. Kjelstrup, J. Sengers), Royal Society Of Chemistry, Cambridge, **2015**, pp. 78–104.

# TOC Elements:

**Keywords**: organic solar cells

**Simulation of bulk-heterojunction stability**: the interplay between liquid-liquid phase separation and crystal growth during ageing of bulk-heterojunctions is investigated by means of phase-field simulations. It is shown that the separated liquid phases guide crystal growth and conversely that crystallites quench the amorphous phase topology. The final film morphology is finer and more structured for higher crystal growth rate.

Olivier J.J. Ronsin * and Prof. Jens Harting *

## Interplay of spinodal decomposition and crystal growth on the stability of crystalline bulk heterojunctions

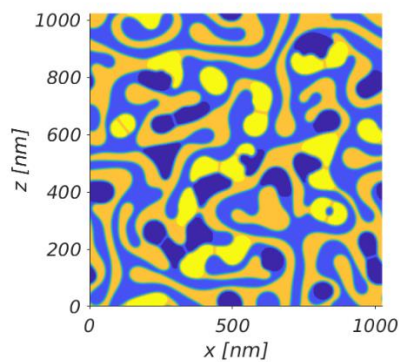